\documentclass[lettersize,journal]{IEEEtran}

\usepackage{amsmath,amsfonts}
\usepackage{algorithmic}
\usepackage{array}
\usepackage[caption=false,font=normalsize,labelfont=sf,textfont=sf]{subfig}
\usepackage{textcomp}
\usepackage{stfloats}
\usepackage{url}
\usepackage{verbatim}
\usepackage{graphicx}
\def\BibTeX{{\rm B\kern-.05em{\sc i\kern-.025em b}\kern-.08em
    T\kern-.1667em\lower.7ex\hbox{E}\kern-.125emX}}
\usepackage{balance}

\usepackage{caption}
\usepackage{graphicx}
\usepackage{tabularx}
\usepackage{ragged2e}
\usepackage{amsmath}
\usepackage[utf8]{inputenc}
\usepackage[english]{babel}
\usepackage{hyperref}
\usepackage{xurl}
\usepackage{wasysym}
\usepackage{multirow}
\usepackage{adjustbox}
\usepackage{array}
\usepackage{booktabs}
\usepackage{xurl}
\usepackage{xspace}
\usepackage{pdflscape}
\usepackage{arydshln}
\usepackage{mathtools}
\usepackage{wrapfig}
\usepackage{float}
\usepackage{csquotes}
\usepackage{xcolor}
\usepackage{acro}

\usepackage{algorithm}
\usepackage{algorithmic}

\usepackage{hyperref}

\usepackage{enumerate}
\usepackage[shortlabels,inline]{enumitem}

\newcolumntype{L}{>{\RaggedRight\arraybackslash}X}
\newcommand{\coin}[1]{\texttt{#1}}
\newcommand{\tokenaddress}[2]{\href{https://etherscan.io/token/#2}{\texttt{#1}}}
\newcommand{\tokenaddressbsc}[2]{\href{https://bscscan.com/token/#2}{\texttt{#1}}}
\newcommand{\tokenaddressftm}[2]{\href{https://ftmscan.com/token//#2}{\texttt{#1}}}

\newcommand{\bi}{\begin{itemize}}
\newcommand{\ei}{\end{itemize}}
\newcommand{\be}{\begin{enumerate}}
\newcommand{\ee}{\end{enumerate}}

\newcommand{\eg}{{e.g., }}

\newcommand{\oldstuff}[1]{}
\newcommand{\info}[1]{}
\newcommand{\old}[1]{}
\newcommand{\optional}[1]{}
\newcommand{\consider}[1]{}
\newcommand{\moved}[1]{}
\newcommand{\comments}[1]{}
\newcommand{\temp}[1]{}

\newcommand{\revision}[1]{{#1}}

\newcommand{\subparagraph}[1]{\par\smallskip\noindent\textbf{#1.}}

\makeatletter
\let\disable\@secondoftwo
\addto\extrasenglish{%
  \renewcommand{\sectionautorefname}{\S\@gobble}%
}
\addto\extrasenglish{%
  \renewcommand{\subsectionautorefname}{\S\@gobble}%
}
\addto\extrasenglish{%
  \renewcommand{\subsubsectionautorefname}{\S\@gobble}%
}
\addto\extrasenglish{%
  \renewcommand{\paragraphautorefname}{\S\@gobble}%
}
\makeatother

\DeclareAcronym{sok}{
  short = SoK,
  long  = systematization of knowledge
}
\DeclareAcronym{defi}{
  short = DeFi,
  long  = decentralized finance
}
\DeclareAcronym{amm}{
  short = AMM,
  long = automated market maker
}
\DeclareAcronym{cfmm}{
  short = CFMM,
  long = constant function market maker
}
\DeclareAcronym{dex}{
  short = DEX,
  long = decentralized exchange,
  short-plural-form = DEXs,
  long-plural-form = decentralized exchanges
}
\DeclareAcronym{lp}{
  short = LP,
  long = liquidity provider,
  short-plural-form = LPs,
  long-plural-form = liquidity providers
}
\DeclareAcronym{pmm}{
  short = PMM,
  long = proactive market maker
}
\DeclareAcronym{dao}{
  short = DAO,
  long = Decentralized Autonomous Organization
}
\DeclareAcronym{nft}{
  short = NFT,
  long = non-fungible token
}
\DeclareAcronym{lmsr}{
  short = LMSR,
  long = logarithmic market scoring rule
}
\DeclareAcronym{ido}{
  short = IDO,
  long = initial DEX offering
}
\DeclareAcronym{ieo}{
  short = IEO,
  long = initial exchange offering
}
\DeclareAcronym{ddos}{
  short = DDoS,
  long = distributed denial-of-service
}
\DeclareAcronym{dns}{
  short = DNS,
  long = domain name server
}
\DeclareAcronym{bgp}{
  short = BGP,
  long = border gateway protocol
}
\DeclareAcronym{mev}{
  short = MEV,
  long = maximal extractable value
}
\DeclareAcronym{bdos}{
  short = BDoS,
  long = blockchain denial-of-service
}
\DeclareAcronym{zkp}{
  short = ZKP,
  long = zero-knowledge proof
}
\DeclareAcronym{mpc}{
  short = MPC,
  long = multiparty computation
}
\DeclareAcronym{l7ddos}{
  short = L7 DDoS,
  long = application-layer distributed denial-of-service
}
\DeclareAcronym{dlt}{
  short = DLT,
  long = distributed ledger technology
}
\DeclareAcronym{bsc}{
  short = BSC,
  long = Binance Smart Chain
}
\DeclareAcronym{dapp}{
  short = dApp,
  long = decentralized application
}
\DeclareAcronym{pbs}{
  short = PBS,
  long = proposer/builder separation
}
\DeclareAcronym{nizk}{
  short = NIZK,
  long = non-interactive zero-knowledge
}
\DeclareAcronym{lbp}{
  short = LBP,
  long = Liquidity Bootstrapping Pool
}
\DeclareAcronym{plf}{
  short = PLF,
  long = protocol for loanable funds,
  long-plural-form = protocols for loanable funds
}
\DeclareAcronym{evm}{
  short = EVM,
  long = Ethereum Virtual Machine
}
\DeclareAcronym{pow}{
  short = PoW,
  long = proof of work
}
\DeclareAcronym{pos}{
  short = PoS,
  long = proof of stake
}
\DeclareAcronym{apy}{
  short = APY,
  long = annual percentage yield
}
\DeclareAcronym{tradfi}{
  short = TradFi,
  long = traditional finance
}
\DeclareAcronym{tvl}{
  short = TVL,
  long = total value locked
}
\DeclareAcronym{iou}{
  short = \revision{IOU},
  long = \revision{``I owe you''}
}

\begin{document}

\title{Reap the Harvest on Blockchain: \\A Survey of Yield Farming Protocols}

\author{Jiahua Xu and Yebo Feng

\thanks{Jiahua Xu is with the Computer Science Department, University College London, UK. Email: jiahua.xu@ucl.ac.uk.

Yebo Feng is the corresponding author. He is with the Computer and Information Science Department, University of Oregon, USA. Email: yebof@uoregon.edu.
}

}


\maketitle

\begin{abstract}

Yield farming represents an immensely popular asset management activity in \acf{defi}. It involves supplying, borrowing, or staking crypto assets to earn an income in forms of transaction fees, interest, or participation rewards at different \ac{defi} marketplaces.
In this systematic survey, we present yield farming protocols as an aggregation-layer constituent of the wider \ac{defi} ecosystem that interact with primitive-layer protocols such as \acfp{dex} and \acfp{plf}. 
We examine the yield farming mechanism by first studying the operations encoded in the yield farming smart contracts, and then performing stylized, parameterized simulations on various yield farming strategies.
We conduct a thorough literature review on related work, and establish a framework for yield farming protocols that takes into account pool structure, accepted token types, and implemented strategies.
Using our framework, we characterize major yield aggregators in the market including Yearn Finance, Beefy, and Badger DAO.
Moreover, we discuss anecdotal attacks against yield aggregators and generalize a number of risks associated with yield farming.

\end{abstract}

\begin{IEEEkeywords}
Decentralized Finance (DeFi), yield farming, yield aggregator, simulation, blockchain
\end{IEEEkeywords}

\section{Introduction}
\label{sec:intro}

\IEEEPARstart{Y}{ield} farming protocols are deemed as the decentralized asset managers on blockchain. After having absorbed crypto assets from users---including both retail and institutional investors, yield farming protocols algorithmically deploy those funds into one or more revenue generating services such as lending and market making. Yield farming protocols have become immensely popular as they seem to create a win-win-win situation: users can earn return on their idle funds through an automated process; yield farming protocols can charge a management fee; other \ac{defi} services can gain more liquidity.

The concept of yield farming was first popularized in mid 2020 by the leading \ac{plf} Compound with the introduction of its governance token \coin{COMP} \cite{Leshner2020CompoundGovernance}. Compound participants get rewarded with newly-minted \coin{COMP} tokens through both lending and borrowing activities, which lead to offsetting some loan costs for borrowers and increasing the return for lenders. 
This incentive scheme was quickly adopted by other protocols such as Uniswap \cite{Uniswap2020Governance} and Yearn Finance \cite{Cronje2020YFIGovernance}) to attract liquidity and participation. As such, on top of the inherently designed benefit that users get for providing liquidity in different kinds of pools (e.g. interest in the case of lending protocols, or fees in the case of providing liquidity in \ac{amm} pools), additional governance tokens are rewarded to users to further encourage their participation in the issuing platform during the early stage of adoption.
The basic yield farming idea was born: the search for opportunities in the \ac{defi} ecosystem to generate returns on otherwise dormant crypto assets. 

\begin{figure}[tp]
\centering   \includegraphics[width=\linewidth]{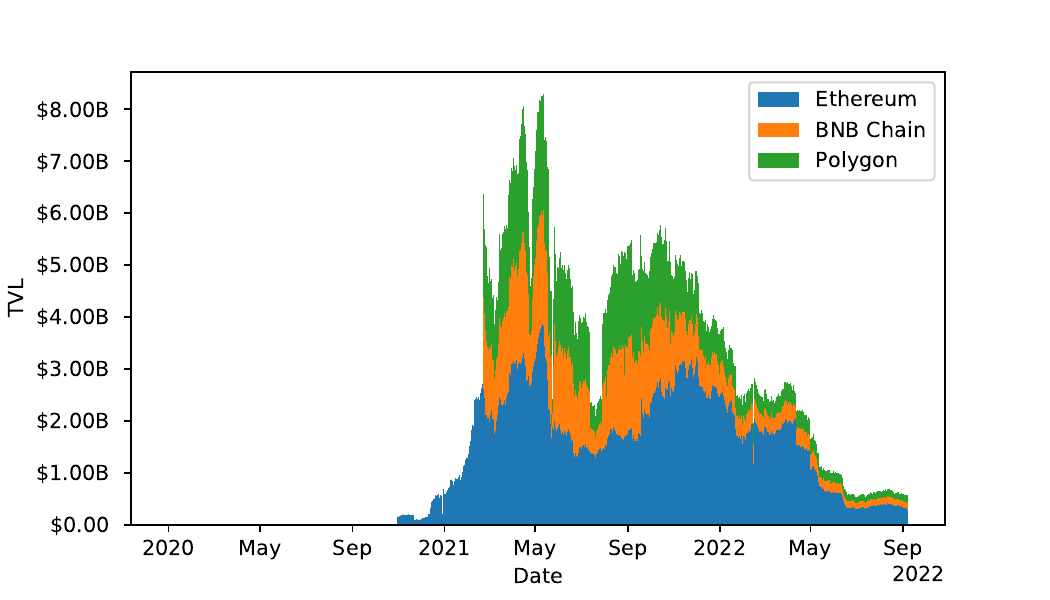}
\caption{\Acf{tvl} (b USD) of yield aggregators on Ethereum, BNB Chain and Polygon. Data collected on 12 September 2022 from \url{https://defillama.com/}.
\label{fig:tvl} 
}
\end{figure}

As a reaction to the creation of a multitude of platforms returning interests, fees and token rewards, yield aggregators---represented by Yearn Finance, Beefy, and Badger DAO (\autoref{tab:top-agg-mkt})---dedicated to farming yield through \ac{defi} primitives emerged. At the beginning 2021, the \ac{tvl} of \ac{defi} yield aggregators was still shy of 1 billion USD; by May 2021, however, this value grew exponentially to 8 billion USD (illustrated in \autoref{fig:tvl}).


In this paper, we present a systemic examination of yield farming protocols.
We first inspect yield farming protocols from the perspective of \ac{defi} architecture and posit them as an aggregation-level component that interact with lower-level primitives in \ac{defi} (see \autoref{sec:background}). We then \revision{synthesize an action-state} framework of yield farming operations, and extract yield farming protocols' features such as pool structure and accepted token types as well as their variations (see \autoref{sec:preliminaries}). With our established model framework, we characterize top yield farming protocols such as Yearn Finance, Harvest Finance and Pickle Finance.
We argue that yield farming protocols are still associated with both security and economic risks (see \autoref{sec:yieldrisks}) and provide a through literature review for interested readers (see \autoref{sec:relatedwork}). 
In Appendix, we present simulations on three typical yield farming strategies in \autoref{sec:formalization}, and describe the workings of top yield aggregators comparatively in \autoref{sec:comparison}. 
Of a particular note here is that this paper is an updated and extended version of work published in~\cite{Cousaert2021}.

\begin{table}[tp]
\tiny
  \centering
  \caption{Top yield aggregators - market share information.}
  \label{tab:top-agg-mkt}
  \setlength{\tabcolsep}{3pt}
    \begin{tabular}{llrrlr}
    \toprule
    \textbf{Yield aggregators} & \textbf{Governance token} & \textbf{\Acs{tvl} (m USD)} & \textbf{MCap (m USD)} & \textbf{Time established} & \textbf{Tokenholders} \\
    \midrule
    \href{https://yearn.finance/\#/portfolio}{Yearn Finance} & \tokenaddress{YFI}{0x0bc529c00C6401aEF6D220BE8C6Ea1667F6Ad93e} & 652.07 & 354.96 & 07/2020 & 49,668 \\
    \href{https://beefy.com/}{Beefy} & \tokenaddress{BIFI}{0x2791BfD60D232150Bff86b39B7146c0eaAA2BA81} & 302.95 & 39.00    & 09/2020 & 25,737 \\
    \href{https://badger.com/}{Badger DAO} & \tokenaddress{BADGER}{0x3472A5A71965499acd81997a54BBA8D852C6E53d}  & 107.22 & 47.83 & 11/2020 & 30,757 \\
    \href{https://idle.finance/}{Idle Finance} & \tokenaddress{IDLE}{0x875773784Af8135eA0ef43b5a374AaD105c5D39e}  & 94.17 & 1.38  & 11/2020 & 3,728 \\
    \href{https://yieldyak.com/}{Yield Yak} & \tokenaddress{YAK}{0x9D5E22B6599c426B453De4a43Df8a0Cb4de061b1}  & 72.13 & 3.56  & 09/2021 & 2,208 \\
    \href{https://autofarm.network/}{Autofarm} & \tokenaddressbsc{AUTO}{0xa184088a740c695e156f91f5cc086a06bb78b827?a=0xa184088a740c695e156f91f5cc086a06bb78b827} & 64.99 & 26.52 & 02/2021 & 65,074 \\
    \href{https://flamincome.finance/}{Flamincome} & \tokenaddress{FLAG}{0x57dD84042ec9507963016596A34EdaD42F7e4CCE } & 59.55 &       & 06/2020 & 43 \\
    \href{https://www.rari.capital/}{Rari Capital} & \tokenaddress{RGT}{0xD291E7a03283640FDc51b121aC401383A46cC623}   & 47.03 & 64.27 & 07/2020 & 6,438 \\
    \href{https://vesper.finance/}{Vesper} & \tokenaddress{VSP}{0x1b40183efb4dd766f11bda7a7c3ad8982e998421}  & 42.62 & 4.35  & 02/2021 & 8,690 \\
    \href{https://www.spool.fi/}{Spool Protocol} & \tokenaddress{SPOOL}{0x40803cea2b2a32bda1be61d3604af6a814e70976} & 39.42 & 3.99  & 12/2021 & 522 \\
    \href{https://harvest.finance/}{Harvest Finance} & \tokenaddress{FARM}{0xa0246c9032bC3A600820415aE600c6388619A14D} & 32.74 & 37.39 & 09/2020 & 13,867 \\
    \href{https://acryptos.com/}{ACryptoS} & \tokenaddressbsc{ACS}{0x4197c6ef3879a08cd51e5560da5064b773aa1d29?a=0x4197c6ef3879a08cd51e5560da5064b773aa1d29} & 30.09 & 2.06  & 12/2020 & 8,078 \\
    \href{https://www.reaper.farm/}{Reaper Farm} & \tokenaddressftm{OATH}{0x21ada0d2ac28c3a5fa3cd2ee30882da8812279b6} & 18.21 & 9.25  & 07/2021 & 2,993 \\
    \href{https://pickle.finance/}{Pickle Finance} & \tokenaddress{PICKLE}{0x429881672B9AE42b8EbA0E26cD9C73711b891Ca5} & 14.20  & 1.49  & 09/2020 & 8,161 \\
    \href{https://onx.finance/}{OnX Finance} & \tokenaddress{ONX}{0xE0aD1806Fd3E7edF6FF52Fdb822432e847411033}  & 4.63  & 1.41  & 03/2021 & 2,941 \\
    \href{https://defiwaterfall.com/}{Waterfall DeFi} & \tokenaddressbsc{WTF}{0xd73f32833b6d5d9c8070c23e599e283a3039823c?a=0xd73f32833b6d5d9c8070c23e599e283a3039823c} & 4.08  & 1.74  & 11/2021 & 852 \\
    \href{https://solidexfinance.com/\#/home}{Solidex} & \tokenaddressftm{SEX}{0xD31Fcd1f7Ba190dBc75354046F6024A9b86014d7} & 3.86  & 0.19  & 02/2022 & 9,389 \\
    \href{https://www.robo-vault.com/}{Robo-Vault} &       & 3.81  &       & 07/2021 &  \\
    \href{https://magik.farm/\#/fantom}{Magik Farm} & \tokenaddressftm{MAGIK}{0x87a5c9b60a3aaf1064006fe64285018e50e0d020} & 3.57  &       & 01/2022 & 2,110 \\
    \bottomrule
    \end{tabular}%
   \label{tab:marketshare}%
   \scriptsize \vskip 1mm
   \RaggedRight Data fetched on 14/08/2022 from  \url{https://defillama.com/} - Yield Aggregators.
\end{table}%

\section{Background in \ac{defi}}
\label{sec:background}

\begin{figure}
\centering
\includegraphics[width=\linewidth]{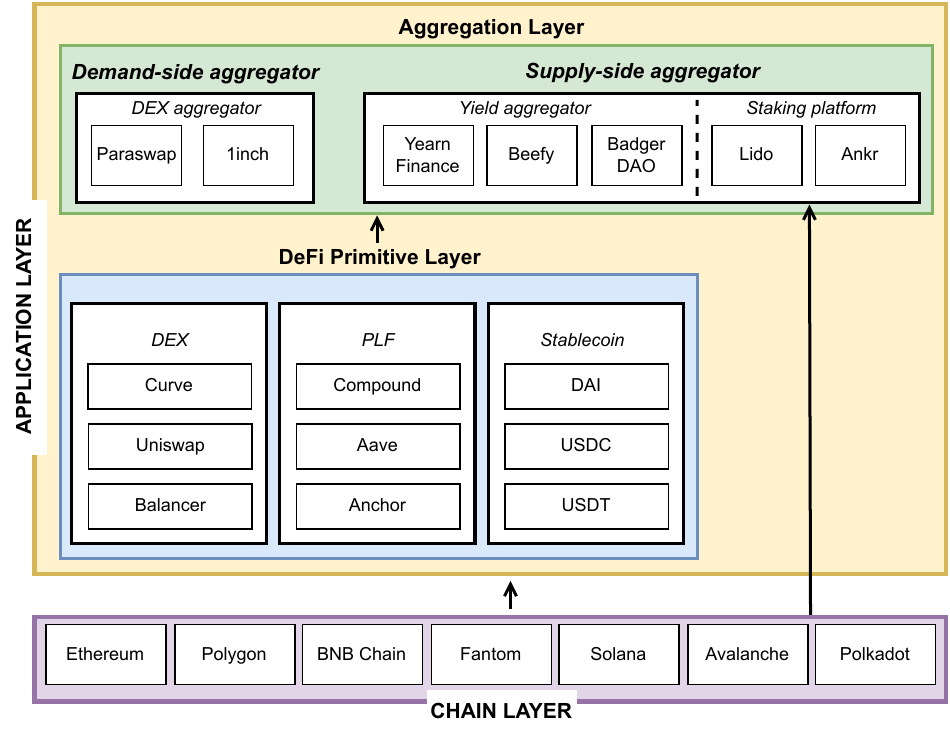}
  \caption{Architecture of the \ac{defi} ecosystem on blockchain.}
  \label{fig:defie_co}
\end{figure}

Yield farming protocols are a constituent part of the wider \ac{defi} ecosystem, and operate heavily dependent on other ecosystem components. In this section, we present those related components to understand where yield farming protocols reside within the \ac{defi} ecosystem (illustrated in \autoref{fig:defie_co}).

\subsection{\revision{\ac{defi} overview}}

\revision{
Built on top of decentralized blockchain networks,
\ac{defi}~\cite{Werner2021sokDefi} systems allow various financial products and services, including lending and asset trading,
to be available to the general public. Compared with traditional financial systems, \ac{defi} democratizes finance by replacing
legacy, centralized institutions with algorithm-backed protocols, thereby improving the accessibility, inclusion, and transparency of financial services~\cite{treleaven2022web,whatsdefi}.}

\subsection{Chain layer}

The \ac{dlt} layer forms the infrastructural basis for \acp{dapp}.
Like all other \acp{dapp},  \ac{defi} protocols consist of one or more smart contracts deployed on blockchain. To this end, the \ac{defi} chain layer typically requires compatibility with smart contracts. 
As the oldest and the most widely adopted \ac{dlt} that supports smart contracts, the Ethereum blockchain is also home to the majority of \ac{defi} protocols \cite{DeFiLeaderboard}. 
The blockchain implements \ac{evm} to ensure that state transitions follow the same rules regardless of node they are performed on.
\revision{The energy consumption and scalability issues associated with blockchains (e.g., EthereumPoW) that are based on the legacy \ac{pow} \cite{Platt2021a} prompted the emergence of the new Ethereum 2.0 and other \ac{evm}-compatible chain layer solutions such as Polygon \cite{polygon2021}, BNB Chain \cite{bnb2022}, Fantom \cite{fantom2018}, and Avalanche \cite{avalanche2020}.} Those solutions incorporate alternative consensus mechanisms like \ac{pos} and exhibit an improved throughput capacity \cite{Perez2020}.
\Ac{pos} chains in particular not only provides the architectural foundation for the \ac{defi} ecosystem, but can also be a source of yield: to encourage users' participation in the consensus of the distributed network, many of these \ac{pos} chains---including Ethereum 2.0 \cite{ethereum2022}, Solana \cite{solana2022} and Polkadot \cite{polkadot2016}---reward users' staking activities.

\subsection{\Ac{defi} primitive layer}

Serving as the fundamental building blocks of the application layer of the \ac{defi} ecosystem, \ac{defi} primitives include \ac{amm}-based \acp{dex}, \acp{plf} and stablecoins. \Ac{defi} yield mainly comes from \ac{amm}-based \acp{dex} and \acp{plf}.

\subsubsection{\Ac{amm}-based \acp{dex}}
\label{sec:amm-dex}

Different from order book-based exchanges where a trade has both buy and sell sides, \ac{amm}-based exchanges---often simply referred to as \ac{amm}---leverage an algorithm termed \enquote{conservation function} to determine the swap rate between two assets given the swap tokens and size \cite{xu2021dexAmm}. As illustrated in Figure~\ref{fig:amm-mech}, traders using an \ac{amm}-based \ac{dex} swap their tokens against the exchange protocol's liquidity pool, which contains tokens deposited by \acp{lp}. 
Against their funds contributed, \acp{lp} receive \enquote{\ac{lp} tokens} as a form of \ac{iou}, which allow liquidity withdrawal and entitle \acp{lp} for their share of swap fee income.
\revision{At the time of writing this paper,} most prominent \ac{amm}-based \acp{dex} include Uniswap \cite{Uniswap2021}, Curve \cite{Curve2021} and Balancer \cite{Balancer2021}. 

\subsubsection{\Acp{plf}}
\label{sec:plf}

A \ac{plf} (illustrated in Figure~\ref{fig:lending-mech}) typically applies a pre-coded interest rate model that dynamically adjusts the borrow and supply rates \cite{Perez2020liquidations}. 
Both rates are commonly programmed to positively correlate with the utilization ratio of the funds, defined as the total amount borrowed as a fraction of the total amount supplied for each specific token asset. 
\acp{plf} on blockchain are mostly collateral-based rather than credit based. This means that a borrow position can only be created when a sufficient amount of deposit is in place acting as collateral. The collateral might become liquidated if market movements or interest accrual cause the borrow position to become insufficiently collateralized.
From the accounting perspective, interest accrual is achieved through \enquote{interest-bearing tokens} which, while sitting in their holder's wallet, increase in value with the passage of time. 
Analogous to \ac{amm}'s \ac{lp} tokens, interest-bearing tokens also serve as a from of \ac{iou}, which are emitted to lenders according to funds supplied and must be surrendered upon funds withdrawal.
Aave \cite{Aave2021} and Compound \cite{Compound2021} can be counted as the two most popular \acp{plf} \revision{at the time of writing this paper}.

\subsubsection{Stablecoins}

Stablecoins are token contracts deployed on blockchain representing cryptocurrencies that offer price stability relative to a certain reference asset \cite{moin2019stablecoin}, namely, a \enquote{peg}. The peg can be another cryptocurrency, legal tender, commodities, or a combination of the above.
\revision{At the time of writing this paper, the biggest stablecoins,} {\tt USDT}, {\tt USDC} and {\tt DAI} are all pegged to the US Dollar. Stablecoins can be custodial or non-custodial, asset-backed or algorithmically programmed. 

\subsection{Aggregation layer}
\label{sec:agg-layer}

\Ac{defi} protocols on the aggregation layer interact with the chain layer or the \ac{defi} primitive layer on end users' behalf \cite{Schar2021Defi}. Depending on whether their target users are requesting or providing services, aggregation layer protocols can be classified as demand-side aggregators and supply-side aggregators. The latter is the category the yield farming protocols belong to. 

\subsubsection{Demand-side aggregators}
\label{sec:demand-side-agg}

Channeling similar services offered by \ac{defi} primitives, demand-side aggregators seek to present users with the most competitive offer so that they do not have to manually perform the comparison themselves. 
\ac{dex} aggregators \revision{Paraswap~\cite{ParaSwap} and 1inch~\cite{1inch}} algorithmically search for the optimal swap route through multiple primitive \acp{dex} to generate the best exchange rate for users.

\subsubsection{Supply-side aggregators}
\label{sec:supply-side-agg}

\begin{figure}[tp]
\centering

\subfloat[\Acf{amm}.\label{fig:amm-mech}]{
\includegraphics[width=\linewidth]{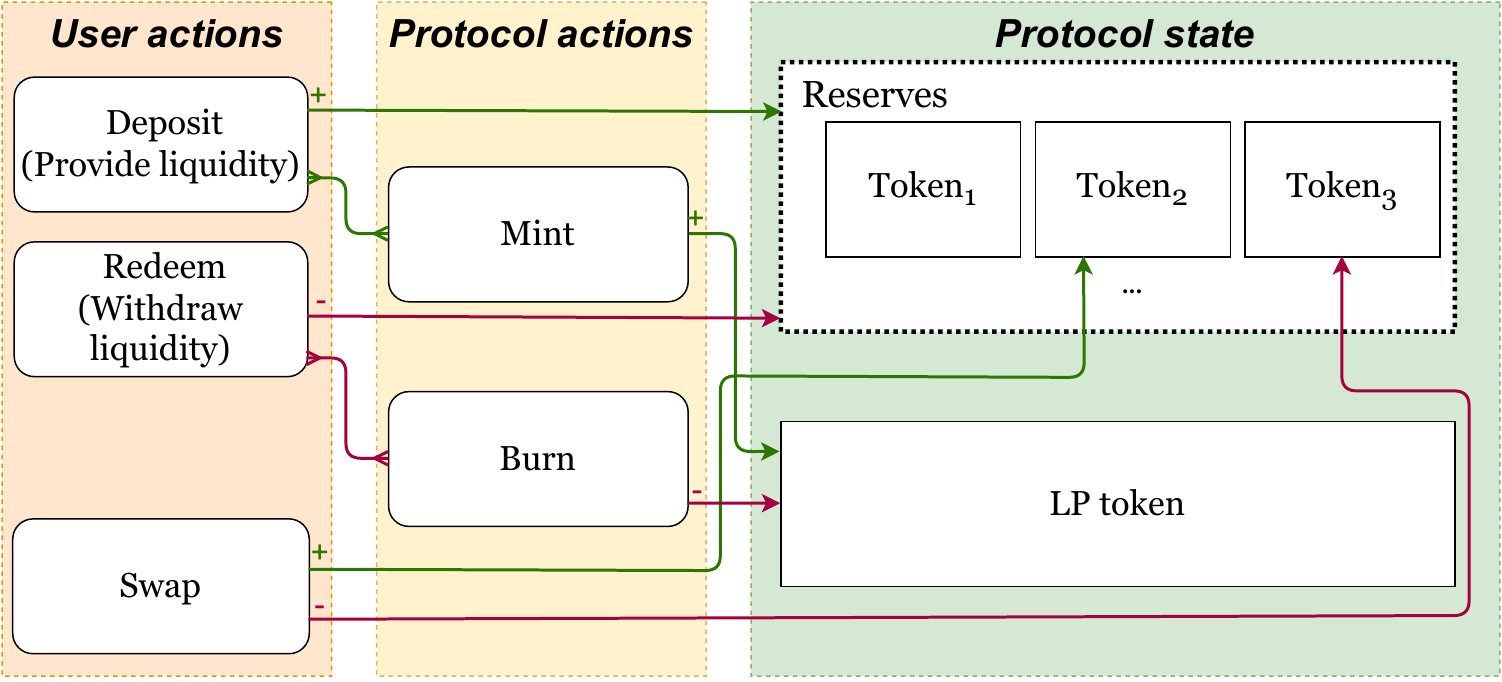}
}

\subfloat[\Acf{plf}.\label{fig:lending-mech}]{
\includegraphics[width=\linewidth]{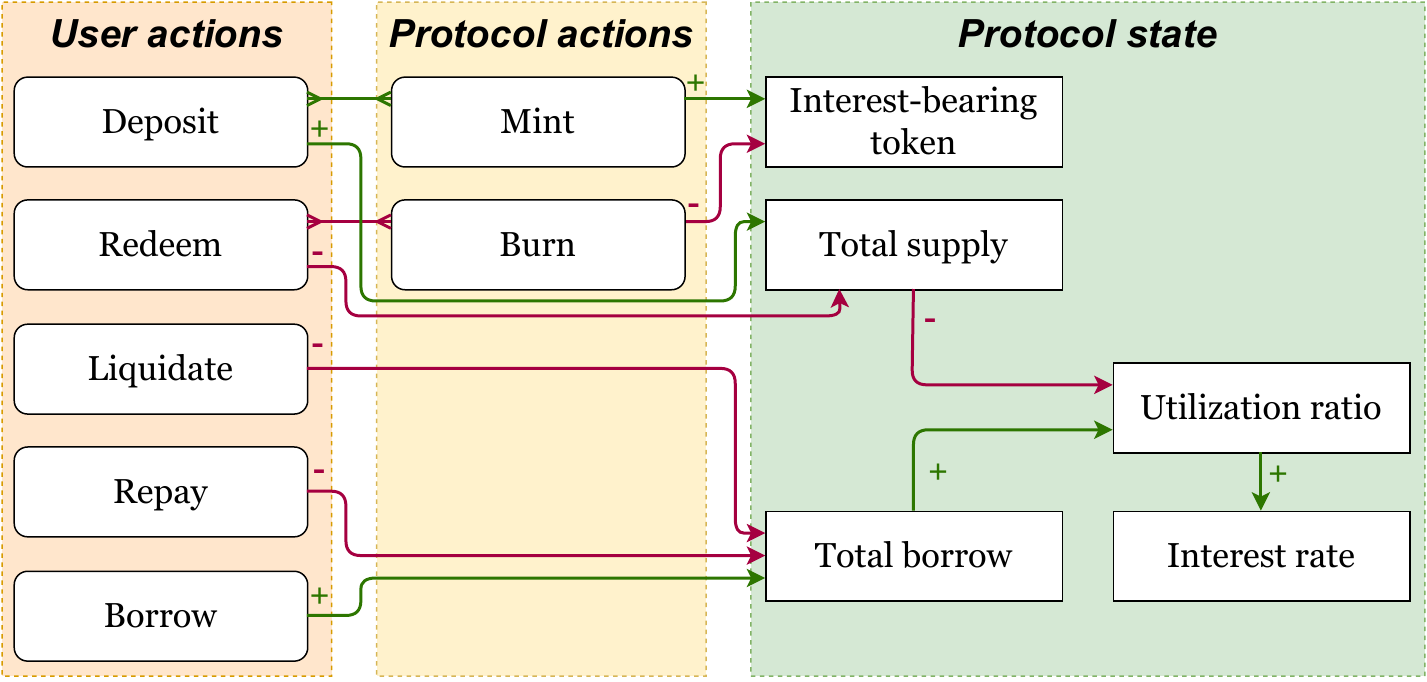}
}

\subfloat[Yield aggregator.\label{fig:yield-mech}]{
\includegraphics[width=\linewidth]{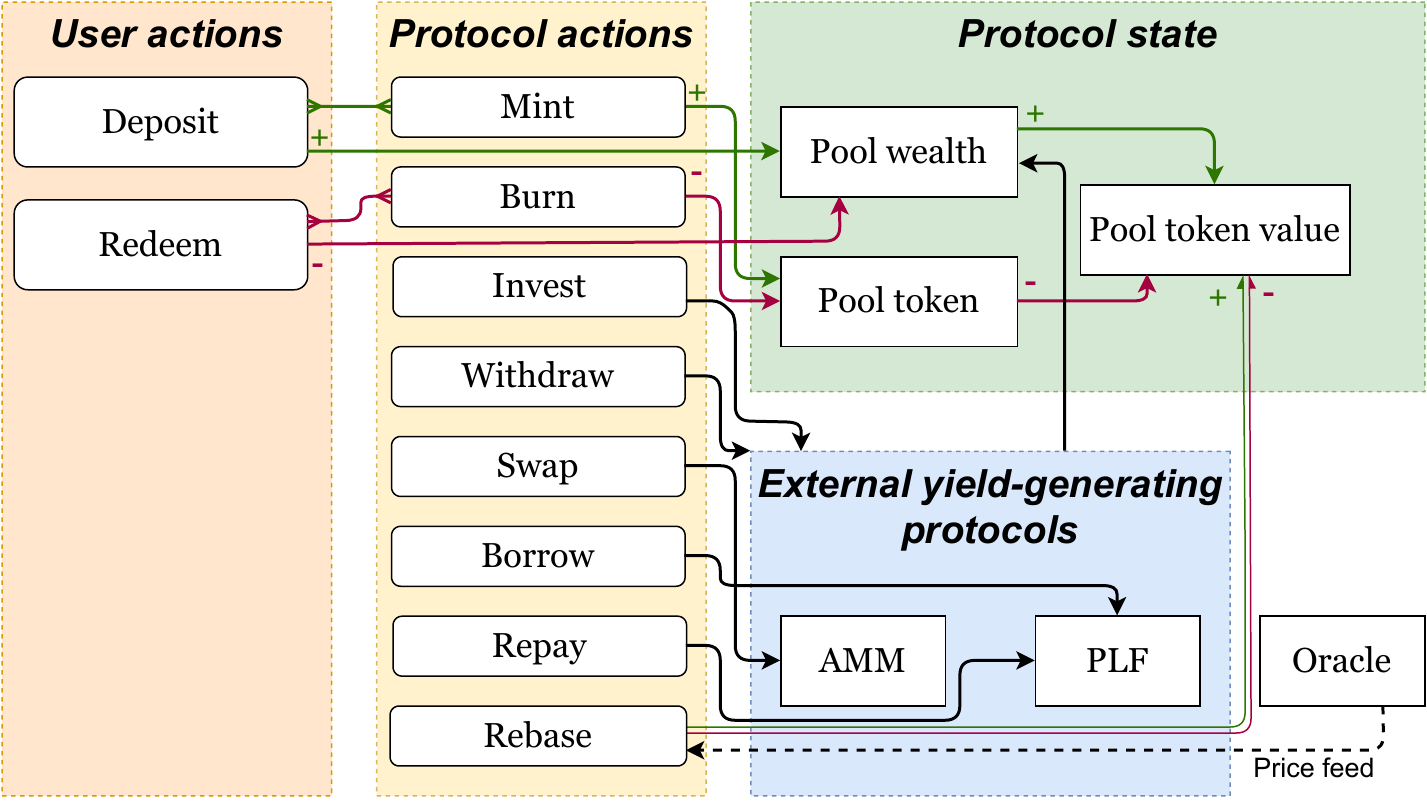}
}

\caption{State diagram with describing the interaction between the environment of a \ac{defi} protocol and associated actions. $+$ means positive effect, $-$ negative effect.}
\label{fig:mech}
\end{figure}

All supply-side aggregators to a certain extent perform some form of yield farming.
Some protocols farm yields directly from the chain layer. For instance, staking platforms like Lido \cite{lido2021home} and Ankr \cite{ankr2022} act as a one-stop shop for users to benefit from staking rewards from various \ac{pos} chains; 
yield aggregators like Yearn Finance \cite{yearn2021}, Beefy \cite{beefydocs} and Badger DAO \cite{badgerdocs} collect users' funds, redeposit them to \ac{defi} primitives such as \acp{dex} and \acp{plf} or other aggregators to generate returns that will be re-distributed back to the users (presented in Figure~\ref{fig:yield-mech}).

\section{Yield farming preliminaries}
\label{sec:preliminaries}

In this section, we dive deep into the workings of yield farming protocols, understand how they generate \revision{yield for the users as well as revenue for the protocols themselves}.

\subsection{Types of yield farming protocols}

There is no universal definition for yield farming protocols. Some equate yield farming protocols to generic yield-generating protocols, in which sense, \ac{defi} primitives such as \acp{amm} and \acp{plf} would also be counted as they offer yield to \acp{lp} and lenders, respectively. 
More commonly, however, yield farming protocols refer to protocols on the aggregation layer (see \autoref{sec:agg-layer}) that pool funds to generate return by interacting with \ac{defi} primitives. 
This is the type of yield farming protocols that we focus on in this paper. 

Besides yield aggregators which are the most widely recognized type of yield farming protocols, some other protocols are more implicit in their farming activities by branding themselves as e.g. stablecoin or lottery protocols. Those protocols mainly differ in the form of \ac{iou} tokens they mint to end users upon new deposit.

\subsubsection{Yield aggregators}
Represented by Yearn, Beefy and Badger, the most classic and commonly known yield farming protocols are yield aggregators. In return for deposit into a yield farming pool, pool tokens that represent a fraction of the pool wealth are issued. Typically, the value of a pool token varies according to the total pool wealth (see \autoref{sec:rebase}). 

\subsubsection{Yield-bearing stablecoins}
A yield-bearing stablecoin protocol works similar to a savings account with a bank.
Instead of minting pool tokens, the protocol issues stablecoins to users as a form of certificate of deposit. 
The yield-generating nature of the protocol is reflected in the increase in the quantity of the stablecoins to their holders, as opposed to the value of the stablecoin token; the value is designed to remain stable to the peg. 
\coin{OUSD} issuer Origin Dollar and \coin{USDi} issuer Bank of Chain \cite{BankofChain2022BankDocs} are two examples of this type of protocols.

\subsubsection{Lottery protocols}

A lottery protocol collect users' funds and issue them each a lottery ticket token in return. The protocol then performs yield farming under the hood.  Instead of distributing yield proportionate to users' deposit, the protocol every once in a while randomly selects one or more winners who can pocket the yield of all participants.
PoolTogether~\cite{pooltogether2021home} is one of the most popular protocols of this type \revision{while writing this paper}.

\subsubsection{\revision{NFT farming}}
\revision{Recently, with the popularity of \acp{nft}, groups began to explore involving \acp{nft} in yield farming. The main goal of \ac{nft} farming is to create liquidity and utility for \acp{nft}, especially in gaming space, thereby earning yields for token owners~\cite{whatsnftfarming}. Axie Infinity~\cite{AxieInfinity}, ZooKeeper~\cite{zookeeper}, Pulsar Farm~\cite{pulsarfarm}, and MOBOX~\cite{mobox} are typical platforms that provide \ac{nft} farming services.}

\subsection{Yield farming operations}
\label{sec:operations}

As illustrated in Figure~\ref{fig:yield-mech}, the entire yield farming process comprises actions from both the user and the protocol sides.  We discuss common action types associated with yield farming protocols. The exact name and implementation of actions may deviate from one protocol to another.

\subsubsection{User actions}
\label{sec:user-actions}

The actions that yield farming users, a.k.a \enquote{farmers}, need to take are often trivial and straightforward.

\paragraph{Deposit}
\label{sec:deposit}

Protocol users simply select their favored yield farming pool and deposit their funds by transferring token assets to the pool smart contract. In return, users receive pool tokens as a form of \ac{iou} which should increase in value with the passage of time due to the yield farmed by the protocol \cite{xu2021dexAmm,Perez2020liquidations}. 

\paragraph{Redeem}
\label{sec:redeem}

Unless there is a timelock, users can redeem their deposited funds plus any yield generated anytime by surrendering their pool tokens.

\subsubsection{Protocol actions}
\label{sec:protocol-actions}

The more sophisticated operations are assumed by the algorithm of the protocol where the actual yield farming is performed automatically under the hood.

\paragraph{Mint}

The protocol mints pool tokens to the user proportionate to the amount of funds deposited, representing their share of the liquidity within the yield farming pool.

\paragraph{Burn}

When a user requests to withdraw funds from a yield farming protocol, pool tokens need to be surrendered by the user and consequently burned by the protocol. 

\paragraph{Invest}

Depending on the \ac{defi} primitive that the yield farming pool interacts with, the yield farming protocol can invest funds collected from users either into an \ac{amm} as a liquidity provider to collect swap fees (see \autoref{sec:liquidityprovision}), or into a \ac{plf} as a lender to earn supply interests (see \autoref{sec:simplelending}). 
A yield farming pool may also invest in another yield farming protocol, often for the benefit of receiving reward tokens.

\paragraph{Withdraw}

When end users request to redeem their funds from a yield farming pool, the pool contract needs to withdraw the corresponding amount of liquidity from the protocol(s) that it has invested in.

\paragraph{Swap}

\enquote{Raw yield} does not always come in the form of the originally deposited assets. Therefore, the yield farming protocol may perform a swap, usually on an \ac{amm}, to convert yield tokens into the same tokens as originally deposited, which are sometimes reinvested to achieve the compounding effect.

\paragraph{Borrow}

Yield farming protocols may use all or part of the funds deposited by users as collateral to borrow from a \ac{plf}. This may need to be performed due to various reasons:
    \begin{enumerate*}[label={(\roman*)}]
    \item to arbitrarily inflate the borrow position to be qualified for more participation reward (see \autoref{sec:leveragedborrow}),
    \item to borrow out assets that can be invested to generate higher yield than the deposited assets.
    \end{enumerate*}

\paragraph{Repay}

Yield farming protocols that take the \enquote{borrow} action may need to partially or fully repay their loans to reduce or close its borrow position if:
    \begin{enumerate*}[label={(\roman*)}]
    \item the borrow position is on the verge of becoming liquidated,
    \item  the collateral must be withdrawn so that it can be invested elsewhere or returned to end users.
    \end{enumerate*}

\paragraph{Rebase}
\label{sec:rebase}

A yield farming pool mints or burns pool tokens depends on the quantity of the asset deposited or withdrawn as well as the exchange rate between the pool token and the asset.
As yield farming progresses, the farming pool usually accumulates wealth and the exchange rate changes. 
Due to diversified investment in various protocols, some yield farming pools may possess an array of assets different from the one deposited by end users. 
Yield farming protocols connect to price oracles to fetch the price of each of these assets, and subsequently calculate the total value held by the pool.
The exchange rate can thus be updated through dividing the latest pool value denominated by the asset deposited by end users with the circulating quantity of the pool tokens. This process of updating the pool token price is termed \enquote{rebase}.

\subsection{Forms of yield farming pools}
\label{sec:form-pool}

Different yield farming protocols vary in terms of their pool structure and token types acceptable by each pool (\autoref{tab:top20yieldagg}). 

\subsubsection{Pool structure}

A yield farming pool may accept deposits in single or multiple assets.

\paragraph{Single asset}
\label{sec:single-asset}

Most yield farming protocols have single-asset pools. While those pools only accept one particular token asset, they may still hold various assets due to different sorts of yield farmed. Typically, those other assets are automatically swapped for the one acceptable as deposits, and reinvested to generate compounded yield (see \autoref{sec:protocol-actions}).

\paragraph{Multiple assets}

A yield farming pool may also accept multiple token assets. Usually assets acceptable by the same pool share a peg. For example, \revision{at the time of writing this paper,} Badger DAO's \coin{ibBTC}/\coin{crvsBTC} pool accept \coin{ibBTC}, \coin{renBTC}, \coin{WBTC} and \coin{ibbtc/sbtcCRV-f}, all pegged to \coin{BTC}.

\subsubsection{Accepted token types}

Yield farming protocols accept various types of tokens, ranging from stablecoins to \ac{lp} tokens.

\paragraph{Stablecoins}

In the recent low-interest environment in the \ac{tradfi} space, yield farming solutions that boast to offer a high single-digit to a double-digit \ac{apy} for USD-pegged stalecoins have been of particular interest. Most yield farming protocols offer stablecoin farming; in fact, among the top 20 yield aggregators, only Badger DAO has no stablecoin pool thus far. 

\paragraph{\Ac{lp} tokens}

Many yield farming pools also accept \ac{lp} tokens. As discussed in \autoref{sec:amm-dex}, \ac{lp} tokens themselves already entitle their tokenholders to swap fee income. Nevertheless, having \ac{lp} tokens managed by a yield farming pool provides the additional benefit of automatically converting and reinvesting participation reward (see \autoref{sec:participation-reward}) distributed by the respective \acp{amm}.

\paragraph{Others}

Other asset types may also be eligible for yield farming. For example, Yearn Finance accepts \coin{ETH}, the native currency on the Ethereum blockchain, as well as \coin{UNI} and \coin{YFI}, which are protocol governance tokens of Uniswap and Yearn Finance itself, respectively.

\begin{table*}[htbp]
  \centering
  \setlength{\tabcolsep}{1.5pt}
  \caption{Top yield aggregators - protocol mechanism.}
  \scriptsize
\begin{tabular}{@{}llccccccccl@{}}
\toprule
          &       & \multicolumn{2}{c}{\textbf{Pool structure}} & \multicolumn{3}{c}{\textbf{Accepted token type}}    & \multicolumn{3}{c}{\textbf{Strategies}}       &        \\
    \cmidrule(lr){3-4}
    \cmidrule(lr){5-7}
    \cmidrule(lr){8-10}
    \textbf{Yield aggregators} &       & {Single asset} & {Multiple assets} & {Stablecoins} & {LP token} & {Others} & {Simple lending} & {Leveraged borrow} & {Liquidity provision} & {Chains} \\

    \midrule
    
    \href{https://yearn.finance/vaults}{Yearn Finance} & \cite{yearndocs} & \CIRCLE & \Circle & \CIRCLE & \CIRCLE & \CIRCLE & \CIRCLE & \CIRCLE & \CIRCLE & Ethereum, Fantom \\
    \href{https://app.beefy.com/}{Beefy} & \cite{beefydocs} & \CIRCLE & \Circle & \CIRCLE & \CIRCLE & \CIRCLE & \Circle & \CIRCLE & \CIRCLE & Polygon, Fantom, BNB \\
    \href{https://app.badger.com/?chain=ethereum}{Badger DAO} & \cite{badgerdocs} & \CIRCLE & \CIRCLE & \Circle & \CIRCLE & \CIRCLE & \CIRCLE & \Circle & \CIRCLE & Ethereum, Fantom, Polygon, BNB \\
    \href{https://app.idle.finance/\#/best}{Idle Finance} & \cite{idledocs} & \CIRCLE & \Circle & \CIRCLE & \CIRCLE & \Circle & \CIRCLE & \Circle & \Circle & Ethereum, Polygon \\
    \href{https://yieldyak.com/farms}{Yield Yak} & \cite{yielddocs} & \CIRCLE & \Circle & \CIRCLE & \CIRCLE & \CIRCLE & \CIRCLE & \CIRCLE & \Circle & Avalanche \\
    \href{https://autofarm.network/}{Autofarm} & \cite{autodocs} & \CIRCLE & \Circle & \CIRCLE & \CIRCLE & \Circle & \CIRCLE & \Circle & \CIRCLE & BNB, Polygon \\
    \href{https://app.flamincome.finance/}{Flamincome} & \cite{flamindocs} & \CIRCLE & \Circle & \CIRCLE & \CIRCLE & \Circle & \Circle & \Circle & \CIRCLE & Ethereum \\
    \href{https://app.rari.capital/}{Rari Capital} & \cite{raridocs} & \CIRCLE & \CIRCLE & \CIRCLE & \Circle & \CIRCLE & \CIRCLE & \Circle & \Circle & Ethereum \\
    \href{https://app.vesper.finance/}{Vesper} & \cite{vesperdocs} & \CIRCLE & \Circle & \CIRCLE & \Circle & \CIRCLE & \CIRCLE & \Circle & \CIRCLE & Ethereum, Avalanche, Polygon \\
    \href{https://www.app.spool.fi/}{Spool Protocol} & \cite{spooldocs} & \CIRCLE & \Circle & \CIRCLE & \Circle & \Circle & \CIRCLE & \CIRCLE & \CIRCLE & Ethereum \\
    \href{https://harvest.finance/}{Harvest Finance} & \cite{harvestdocs} & \CIRCLE & \CIRCLE & \CIRCLE & \Circle & \CIRCLE & \CIRCLE & \Circle & \CIRCLE & Ethereum, Polygon, BNB \\
    \href{https://app.acryptos.com/}{ACryptoS} & \cite{acsdocs} & \CIRCLE & \Circle & \CIRCLE & \CIRCLE & \CIRCLE & \Circle & \Circle & \CIRCLE & BNB, Fantom \\
    \href{https://www.reaper.farm/}{Reaper Farm} & \cite{reaperdocs} & \CIRCLE & \CIRCLE & \CIRCLE & \CIRCLE & \CIRCLE & \Circle & \Circle & \CIRCLE & Fantom \\
    \href{https://app.pickle.finance/}{Pickle Finance} & \cite{pickledocs} & \CIRCLE & \Circle & \CIRCLE & \CIRCLE & \CIRCLE & \Circle & \Circle & \CIRCLE & Ethereum, Polygon \\
    \href{https://app.onx.finance/vaults?platform=All&type=All&asset=All&sort=DEFAULT&reward=All&isDeprecated=false}{OnX Finance} & \cite{onxdocs} & \CIRCLE & \Circle & \CIRCLE & \CIRCLE & \CIRCLE & \Circle & \Circle & \CIRCLE & Ethereum, Polygon, Fantom, Avalanche \\
    \href{https://app.defiwaterfall.com/}{Waterfall DeFi} & \cite{waterfalldocs} & \CIRCLE & \CIRCLE & \CIRCLE & \CIRCLE & \CIRCLE & \CIRCLE & \CIRCLE & \CIRCLE & Avalanche, BNB \\
    \href{https://solidexfinance.com/\#/pools}{Solidex} & \cite{solidexdocs} & \CIRCLE & \Circle & \CIRCLE & \Circle & \CIRCLE & \Circle & \Circle & \CIRCLE & Fantom \\
    \href{https://www.robo-vault.com/vaults}{Robo-Vault} & \cite{robodocs} & \CIRCLE & \Circle & \CIRCLE & \CIRCLE & \Circle & \CIRCLE & \Circle & \CIRCLE & Fantom, Avalanche, Polygon \\
    \href{https://magik.farm/\#/fantom}{Magik Farm} & \cite{magikdocs} & \CIRCLE & \Circle & \CIRCLE & \CIRCLE & \CIRCLE & \Circle & \Circle & \CIRCLE & BNB, Avalanche, Fantom \\
    \bottomrule
    \end{tabular}%
  \label{tab:top20yieldagg}%
  \scriptsize \vskip 1mm
  \RaggedRight Data fetched on 28/08/2022 from corresponding documents.
\end{table*}%

\subsection{Sources of yield}

\subsubsection{Supply interest}
The most straightforward type of yield originates from lending. 
As the demand for loans in crypto assets grows, the borrowing interest rate increases, leading to higher yields for lenders.
Particularly in a bullish market, speculators are keen to borrow funds despite a high interest rate, in expectation of an appreciation in the assets of their leveraged long position.
A borrower wishing to increase their exposure to \coin{ETH}, for example, may use \coin{ETH} as collateral to borrow \coin{USDC}, then repetitively exchanging \coin{USDC} for \coin{ETH} to deposit it as collateral to borrow more \coin{USDC}, forming a \enquote{leveraging spiral} \cite{Xu2021c}.
Compound \cite{Compound2021} and Aave \cite{Aave2021}, \revision{two major DeFi lending protocols while writing this paper} (see \autoref{sec:plf}), have witnessed the borrow APY of \coin{USDC} rising from 2-3\% in May 2020 to as high as 10\% in April 2021.\footnote{\url{https://app.defiscore.io/assets/usdc}} 
This specific kind of yield is incorporated in interest-bearing tokens, such as \coin{cTokens} from Compound or \coin{aTokens} from Aave.

\subsubsection{Swap fee income}
Some tokens entitle users to part of the revenue that is going through the protocol. These can be governance tokens or other kinds of tokens. 
One example is the liquidity provider tokens in AMM-based DEXs \cite{xu2021dexAmm}. By supplying liquidity into an AMM pool, users receive the fees that are paid by traders within that pool. The higher the volume in that pool, the more fees that are generated, and the more a liquidity provider profits from this. In Uniswap \cite{Uniswap2021}, a 0.3\% fee is charged for every trade within a pool and goes fully to LPs.

\subsubsection{Participation reward}
\label{sec:participation-reward}

Another yield source comes from liquidity mining programs, where early participants receive native tokens representing protocol ownership. This incentivizes people to contribute funds into the protocol, and enhances decentralization as the protocol ownership is distributed to users. 
The native tokens often have a governance functionality attached to them which is deemed valuable, as the token holders have a say in the future strategic direction of the project. Native tokens sometimes also entitle holders to a share of the protocol revenue. Further, the values these tokens possess itself especially in a speculation context can be the benefits of owing a protocol.

This brings up a second kind of revenue-sharing token, where users have to actively stake their tokens to receive a share of the revenue. For example, \coin{SUSHI} holders that stake their \coin{SUSHI} will get \coin{xSUSHI} in return, which represents the proportional share of a pool that captures 0.05\% of all trades on Sushiswap \cite{xsushi2021faq}. Vesper Finance's governance token, \coin{VSP}, can also be deposited in a pool, in return for \coin{vVSP}, a token that represents the user's proportional share of a pool that captures part of the revenue generated throughout the whole Vesper platform~\cite{vesper2021tokenomics}.

\subsection{Revenue model of yield farming protocols}

Yield farming protocols often retain a fraction of yield earned as the protocol revenue \cite{Xu2022b}. In the spirit of \ac{dao}, the revenue may be redistributed to tokenholders of the protocol governance token \cite{Xu2022e}. In that sense, a stronger buy pressure of a particular governance token of a yield farming protocol usually mirrors a larger (anticipated) \ac{tvl} of the protocol, as it can translate to higher protocol revenue.

\section{Yield farming risks}
\label{sec:yieldrisks}

Compared with asset management in \acf{tradfi}, yield farming may bring substantial profits in short order but also carry a range of risks.

\subsection{Security risks}

We identify four major types of attacks associated with yield farming. \autoref{tab:attack} presents anecdotal events of these attacks and their potential solutions.

\subsubsection{Flash loan attack}

Flash loan attacks~\cite{Qin2020c} abuse the mechanism of flash loan protocols in which an attacker borrows a great deal of funds that do not require collateral. The attacker then manipulates the price of an asset in a very short period and quickly resell it to earn profits. Such a procedure can be repeated for multiple time by the attacker, thereby causing considerable losses to investors and yield aggregators.

Major yield aggregators have witnessed multiple waves of flash loan attacks, losing millions of dollars each. For example, ApeRocket suffered two flash loan attacks that costed investors 1.26 million USD on July 2021~\cite{Aperocket_c1}; in October 2020, a farmer leveraged flash loans to reap 33.8 million USD from the \coin{USDT} and \coin{USDC} pools~\cite{harvest_c1}; Pancake Bunny Finance lost around 690,000 \coin{BUNNY} tokens due to removal of liquidity and price manipulation by flash loan attacks~\cite{Bunny_c1}. 

To defend against flash loan attacks, developers must consolidate and improve flash loan protocols, making them difficult to be exploited by attackers. An effective approach is to setup floating interest rates, thereby increasing the cost of launching flash loan attacks~\cite{Bunny_c1}. Developers can also choose to enhance the audits~\cite{ApeRocket_c3} or forbid depositing and withdrawing funds within a single transaction~\cite{harvest_c3}.

\subsubsection{Rug pull}
A rug pull refers to the abandonment of a project by the project administrator after collecting investor’s funds, leaving investors with valueless assets~\cite{xu2021dexAmm,rug-pull-businessreview}. One way of conducting this type of scam is to lure yield farming protocols into buying assets with no value and then swap this asset for \coin{ETH} or another type of asset with value. For example, Arbix Finance, a typical rug pull, drained around 10 m USD in users' assets directly from the vaults without any advanced attack techniques in sight~\cite{arbix1}.

To prevent from being rugged, investors should exercise caution and always confirm a project's credibility before investing in it~\cite{xia2021trade,mackenzie2022criminology}. Besides, continuously tracking the audit information of invested projects enables investors to quickly identify risks and take appropriate measures, thereby reducing losses~\cite{arbix2}.

\subsubsection{Reentrancy attack}

Even though the composability factor of DeFi is what makes yield farming possible in the first place by allowing for complex, interconnected financial protocols, it does bring along the danger of smart contract risk as more and more money legos are plugged into a strategy. While two smart contracts may be secure in isolation, the combination of them may not. By composing multiple smart contracts together, the attack surface might be greater than the sum of its parts~\cite{wang2021ethereum,wan2021smart}.

Reentrancy attack is one of the most destructive attacks that appear when multiple smart contracts operate with each other. More specifically, the reentrancy attack occurs when a smart contract makes an external call to another smart contract. Then the another contract makes recursive calls back to the original function, intentionally or unintentionally withdraw funds. When the original contract fails to update its state before sending funds, the attacker can exploit this vulnerability to continuously drain the contract’s funds.

Major yield aggregators have witnessed a large amount of reentrancy attacks in the past several years. In April 2021, the ForceDAO DeFi aggregator was exploited by a group of attackers, who utilized reentrancy attacks to steal 367 thousands USD worth of tokens before the ForceDAO team took effective actions to prevent further attacks~\cite{badger3}; in September 2021, DAO Maker, a decentralized finance platform on Ethereum, was hacked for almost 4 million USD due to insecure smart contracts~\cite{daomaker1}; a reentrancy attack on the Grim Finance project within the Fantom Blockchain also successfully drained over 30 million USD worth of tokens in 2021~\cite{grim1}.

To defend protocols against reentrancy attacks, researchers and developers have proposed a variety of frameworks and methods~\cite{huang2019smart}.
For example, Rodler et al.~\cite{Rodler2019a} propose a backward compatible approach based on run-time monitoring and validation to protect smart contracts on Ethereum;
Das et al.~\cite{das2021resource} propose a reentrancy-aware language called Nomos, which enforces reentrancy security using resource-aware session types;
Cecchetti et al.~\cite{cecchetti2021compositional} first formalize the reentrancy interface on general distributed systems and then leverage information flow control to automatically fix defective smart contracts.
However, with the increasing complexity and variety of \ac{defi} protocols, reentrancy attacks will also become increasingly difficult to detect and counter. From users' side, protection can be sought from \ac{defi}-native insurance protocols such as Nexus Mutual that cover smart contract risks \cite{Cousaert2022}. 

\subsubsection{Key exploit}

Due to poor access control of some DeFi systems, yield aggregators can be attacked by exploiting various keys (e.g., API key, wallet key) to tamper with the smart contracts or drain funds. For example, the Bent Finance utilized non-multisig wallets to deploy their project’s smart contracts. Anyone who knows the appropriate private key can perform updates to the contracts, allowing attackers to inject malicious code and create the backdoor. In December 2021, an attacker leveraged this feature to drain 1.75 million USD worth of tokens from the pool~\cite{bent1, bent2, bent3}.

To avoid attacks based on key exploiting, DeFi contracts should always be deployed upon multisig wallets to eliminate single points of failure~\cite{bent3}. DeFi platforms should also properly protect the private keys used to access and control correlative smart contracts. The developments of DeFi system API, application, and user interface should follow software security practices, ensuring that the access control and function calls are solidly implemented.

\subsubsection{Other attacks}
As yield farming protocols are built upon multiple complex systems with a variety of software and hardware components interacting with each other, both technical and economic weaknesses give rise to attractive exploit opportunities for malicious hackers. Besides the aforementioned attacks, there are many other attacks targeting the blockchain infrastructure, user interface, or even network communications, thereby disturbing the proper operation of yield farming protocols. For example, malicious miners can prioritize transactions in their favor by inspecting \acp{mev}, thereby causing damages to the smart contracts running on the upper layers~\cite{qin2022quantifying, zhou2021just}; attackers can break the network connections between the users and the blockchain system through border gateway protocol (BGP) hijacking~\cite{apostolaki2017hijacking}; malicious traders can leverage front-running attacks to drain funds from pools ~\cite{eskandari2019sok}.

These attacks are out of scope for this paper, but it is important for users and developers to be aware of that yield farming security is a systemic problem. Only by ensuring the security of every component in the system can the security of yield seekers' funds be ensured.

\begin{table*}[tb]
  \centering
  \caption{Overview of attacks in aggregators and potential solutions.}
  \label{tab:attack}
  \tiny
  \renewcommand{\arraystretch}{1.2}
    \begin{tabularx}{\linewidth}{llXXrlll}
    \toprule
    \textbf{Attacks} & \textbf{Yield aggregator attacked} & \textbf{Summary} & \textbf{Solutions} & \textbf{Estimated lost } & \textbf{References} & \textbf{Time} & \textbf{Major chain}
    \\
\midrule
    Flash loan attack & ApeRocket & Using the fact that the AutoCake vault was only deployed 10 hours and was low in TVL, attacker conducted price manipulation and drained the vault. & Project team updated the protocol and at least two audits will be conducted before its V2 launch. & 1.26 m USD & \cite{Aperocket_c1, ApeRocket_c2, ApeRocket_c3} & 07/14/2021 & BNB\\

          & Pancake Bunny & Within the timeframe to create a new block, attacker transferred USDT into the contract and called removal of liquidity. Caused the value of Bunny token to crash by more than 95\%. & A implementation of the Floating Rate of Emissions and the security code changes. & 3 m USD & \cite{Bunny_c1, Bunny_c2, Bunny_c3,Bunny_c4, Bunny_c5} & 05/20/2021 & BNB \\

          & Harvest Finance & Attacker swapped USDC to USDT to up the price of USDT, depositing USDT into vault and swap back USDT to USDC to gain profit as USDT price fall. This action is repeated to drain the vault. & Team updated the following: deposit and withdraw funds within a single transaction is not allowed to avoid flash loan, and withdraw of tokens are made into multiple transactions to minimize damage. & 33.8 m USD & \cite{harvest_c1, harvest_c2, harvest_c3}  & 26/10/2020 & Ethereum \\
          
\midrule        
    Rug pull & Arbix Finance & The project team drained the vault with users assets, deleted their website, twitter and telegram. & Certik sent out a community alert. & 10 m USD & \cite{arbix1, arbix2} & 01/04/2022 & BNB \\
\midrule 
    Reentrancy attack & ForceDAO & The xFORCE platform used a fork of xSUSHI contract which revert the token if transaction fails, they also used Aragon Minime token that return false if a transferFrom() call fails. & Team could have used a standard Open Zeppelin ERC-20 or added a safe transferFrom wrapper in xSUSHI contract. & 367 k USD & \cite{forcedao_c1, forcedao_c2} & 04/03/2021 & BNB\\

          & Grim finance & Attacker exploited a depositFor() function that had not been protected. Users deposited funds in to vaults that attacker inserted their own contract containing the reentrancy deposit loops. & The team updated the code and send in for an audit. & 30 m USD & \cite{grim1, grim2, grim3, grim4} & 19/12/2021 & Fantom  \\
    
          & DAO Maker & The init() function was vulnerable, attacker initialized 4 token contracts with malicious data then used the emergencyExit() function to drain funds. & The source code is not public so protecting and checking the project is a priority. Also to fix the vulnerability in the function. & 4 m USD & \cite{daomaker1, daomaker2} & 09/03/2021 & Ethereum \\

          & Reaper Farm & Attacker took advantage of that the recipients account verification had not been set up properly and drained the vault. & The project team closed down the vaults attacked, altered the code and waiting for full audit before launching again. & 1.7 m USD & \cite{reaper1, reaper2, reaper3} & 01/08/2022 & Fantom  \\
          
\midrule         
    Key exploit & Bent Finance & The contract used a non multisig wallet, allowing anyone that knows the private key to modify updates, which caused the attacker to create a back door. Attacker altered the code so that Bent finance would provide large amount of funds to the attacker's address. & Project team could have used multisig wallet to avoid and protected private keys in an appropriate way. & 1.75 m USD & \cite{bent1, bent2, bent3} & 12/21/2021 &  BNB \\

        & Badger DAO & Attacker used a compromised API key to periodically inject malicious code into the contract. These codes are triggered when users try to perform transactions, allowing unlimited spend approvals for the attacker’s address. & Project team working with cybersecurity firm to fix the problem, as well as authorities to recover any funds possible. & 120 m USD & \cite{badger1, badger2, badger3} & 02/12/2021 & BNB \\
    \bottomrule    
    \end{tabularx}
\end{table*}%

\subsection{Economic risks}

Besides security concerns, there exist various economic risks associated with yield farming. 
In \autoref{sec:sim-results}, we demonstrate that investment strategies with the potential to generate remarkably high yield also bear high risks. While our simulation only illustrate return courses in a deterministic fashion, through various simulated scenarios one can easily extrapolate that
the ever-changing market conditions---including volatile price movements and trading activities---lead to return instability, and sometimes even losses.
Below, we discuss several types of economic risks associated with yield farming.

\subsubsection{Yield dilution risk}

Yield farming pools providing double or even triple digit \ac{apy} can be deceiving in their return generating capability. Often enough, those pools are thin in liquidity and lack scalability, unable to accommodate a large amount of deposit while sustaining a similar level of \ac{apy}. Users who invest a significant amount of funds into such a pool may find the pool \ac{apy} dropping significantly immediately afterwards. For users with funds already in the pool, they may experience a decrease in return on their investment due to dilution from newly added funds.
For those who do not constantly monitor their investment performance, this may mean leaving their funds in a diluted, low-\ac{apy} pool, while missing the opportunity of reallocating their funds to more profitable strategies.

\subsubsection{Conversion risk}

As discussed in \autoref{sec:form-pool}, yield farming pools typically specify tokens that they can accept. Therefore, to participate in yield farming, users may have to first convert partially or all of their funds into acceptable assets for yield farming.
This engenders conversion risk: a user might have been better off holding their original funds, than converting them to \enquote{eligible} assets. This is because the \enquote{eligible} assets might depreciate against the original assets prior to the conversion to such an extent that even the yield generated cannot make up for the depreciation loss.
This risk is most prominent with liquidity provision strategies, manifested by the so-called \enquote{impermanent loss} (see \cite{xu2021dexAmm}). By design, the value \ac{lp} tokens (e.g. \coin{USDT-ETH-LP} token) from an \ac{amm}-based \ac{dex} falls against the original portfolio (e.g. a combination of \coin{USDT} and \coin{ETH}). Sometimes, even the swap fee income and the participation reward are insufficient to cover the conversion loss.

\subsubsection{Exchange risk}

Related to conversion risk, exchange risk is associated with the uncertainty surrounding the exchange rate between the assets held by the yield farming pool and the denominating currency (usually USD).
As demonstrated in \autoref{sec:sim-results}, yield farming strategies benefit from---and, in cases such as leveraged borrow, solely rely on---the high value of participation reward. This makes yield farming highly speculative as token prices are unpredictable. An overly low price of yielded tokens such as reward tokens might result in a loss for end users.

\subsubsection{Counterparty risk}

This risk is associated with farming strategies that incorporate lending, where loans might not be repaid.
While the simple lending strategy (see \autoref{sec:simplelending}) is a relatively low-risk one, losses may still occur under extreme market conditions, e.g. when the price of the asset lent out relative to the collateral suddenly increases to such a significant extent that the loan becomes undercollateralized (see lending protocol MakerDAO's Black Thursday Incident \cite{Kjaer2021,Perez2020liquidations}). 
In such cases, borrowers may choose not to repay their loans since their collateral is not worth the effort anymore, resulting in a default.
Kao et al.~\cite{Kao2020a} simulate a wide range of market volatility to stress-test lending protocols such as Compound, and find that only rarely can undercollateralization occur.

\subsubsection{Liquidation risk}

Liquidation risk is associated with farming strategies such as leveraged borrow (see \autoref{sec:leveragedborrow}) that incorporate taking a overcollateralized loan on a \ac{plf}. Due to price movements and interest accrual, a loan position may become insufficiently collateralized, triggering liquidation of the deposited assets backing the loan. At liquidation, the value of the collateral liquidated by design exceeds the loan payable reduced, resulting a loss on the side of the borrower. Thus, yield farming protocols that implement borrow but unable to handle liquidation risk properly may cause users to lose their funds.

\begin{table*}[t]
  \setlength{\tabcolsep}{1pt}
  \caption{Overview of related literature.}
  \renewcommand{\arraystretch}{1.2}
  \begin{adjustbox}{width=\textwidth}
  \centering
  \scriptsize
  \begin{tabularx}{\linewidth}{lXccccccccccccc}
    \toprule
&  & \multicolumn{3}{c}{\textbf{Subjects Covered}} & \multicolumn{4}{c}{\textbf{Methodology}}                                               \\
    \cmidrule(lr){3-5}
    \cmidrule(lr){6-9}
        \textbf{Ref.} & \textbf{Summary} & {Yield}   & {Major} & {Risk} & {Literature}  & {Modeling} & {Empirical} & {Taxonomization} \\
        &                     & {farming}      & {yield} & {evaluation} & {review} & & {analytics} \\
        &                     & {strategies}      & {aggregators} &  & & &  \\

    \midrule
    This paper & A survey that uses examples of yield strategies to compare the major yield aggregators, translating them into revenue models and empirical examinations supported with on-chain data. & \CIRCLE & \CIRCLE & \CIRCLE & \CIRCLE & \CIRCLE & \CIRCLE & \CIRCLE \\
    \revision{\cite{augustin2022yield}} & \revision{A paper that characterizes the risk and return characteristics of yield farming investment strategies on PancakeSwap, one of the largest automated market makers among the emerging ecosystem of decentralized financial services.} & \CIRCLE & \LEFTcircle & \CIRCLE & \LEFTcircle & \CIRCLE & \CIRCLE & \Circle \\
    \cite{Saengchote2021stablecoins} & A presentation of yield-chasing behavior adopting on-chain transaction level data and empirical analysis to support the result. The paper also classifies DeFi protocols including major yield aggregators. & \CIRCLE & \CIRCLE & \Circle & \Circle & \Circle & \CIRCLE & \CIRCLE \\
    \cite{Walton2022} & A break-down of the mechanism of yield generating, also it provides an overview of four different strategies and other related \ac{defi} services. & \CIRCLE & \CIRCLE & \CIRCLE & \Circle & \Circle & \Circle & \CIRCLE \\   
    \cite{Kjaer2021} & A case study assessing the stability of MakerDAO protocol, which uses public data and protocol analysis. & \CIRCLE & \Circle & \CIRCLE & \Circle & \Circle & \CIRCLE & \Circle \\
    \cite{Saengchote2022compound} & A case study about Compound with detailed explanation of how it works, and where it is applied.  & \CIRCLE & \CIRCLE & \CIRCLE & \Circle & \Circle & \CIRCLE & \LEFTcircle \\
    \cite{han2022} & An analysis of impact on trading using data from Binance's and Uniswap's program in yield farming. It also compares their differences and evaluate how DeFi provides an alternative solution to the traditional finance. & \CIRCLE & \Circle & \CIRCLE & \Circle & \Circle & \Circle & \CIRCLE \\ 
    \cite{Dragos2020} & A discussion of the  transition from the traditional finance to DeFi and its advantages, covering methods of yield farming.  & \CIRCLE & \Circle & \CIRCLE & \Circle & \Circle & \Circle & \CIRCLE \\
    \cite{Xu2021c} & A presentation of the advantageous DeFi characteristics that would resolve a list of fundamental issues in the traditional lending system. & \CIRCLE & \CIRCLE & \CIRCLE & \Circle & \Circle & \Circle & \CIRCLE \\
    \cite{Bartoletti2020sokLendingPools} & A systematic analysis on lending pools and their behavior, focuses on two specific lending platforms.  & \CIRCLE & \Circle & \CIRCLE & \Circle & \CIRCLE & \Circle & \LEFTcircle \\
    \cite{Gudgeon2020PLF} & A discussion of the Protocol for Loanable Funds, also reviews the methodology of interest rate determination and provides empirical examination corresponding to different degrees of liquidation. & \CIRCLE & \Circle & \CIRCLE & \Circle & \CIRCLE & \Circle & \CIRCLE \\
    \cite{Perez2020liquidations} & An empirical analysis of liquidation on Protocol for Loanable Funds, focuses on Compound. The paper also provides calculation of liquidators efficiency and discusses the security issues and risks. & \CIRCLE & \Circle & \CIRCLE & \Circle & \CIRCLE & \CIRCLE & \Circle \\
    \cite{moin2019stablecoin} & A discussion on the structure of stablecoins that breaks down existing stablecoins into components to compare and evaluates their advantages and disadvantages. & \CIRCLE & \Circle & \Circle & \Circle & \Circle & \Circle & \LEFTcircle \\
    \cite{xu2021dexAmm} & An SoK on AMM based DEX protocols, compares the popular protocols mechanisms and discusses the securities and privacy concerns. & \CIRCLE & \Circle & \CIRCLE & \CIRCLE & \CIRCLE & \CIRCLE & \CIRCLE \\
    \cite{1952Markowitz} & A discussion on the method of maximizing the discounted value of future returns and analyzes the geometric relationships between the expectation and the choice of portfolio. & \Circle & \Circle & \CIRCLE & \Circle & \CIRCLE & \Circle & \LEFTcircle \\
    \bottomrule

    \end{tabularx}%
    \end{adjustbox}
  \label{tab:related_work}%
\end{table*}%
\section{Related work}
\label{sec:relatedwork}

In this section, we introduce literature that is related to yield farming in some shape and form. As it is still a fairly new area, there is a paucity of existing works related to our paper. \autoref{tab:related_work} summarizes the most related and representative ones.

In general, our paper is different from existing papers in the following aspects:

\begin{itemize}
    \item our paper focuses on the subject of yield farming, while other papers either investigate a more specific \ac{defi} topic (\eg yield generating mechanism~\cite{Walton2022}, yield chasing~\cite{Saengchote2021stablecoins}), examine a different but related DeFi applications (\eg AMM-based \acp{dex}~\cite{xu2021dexAmm}, lending~\cite{Bartoletti2020sokLendingPools}), or have a broader coverage (\eg DeFi transformation~\cite{Dragos2020});
    \item to investigate yield farming comprehensively, our paper utilizes multiple methodologies, including literature survey, modeling, empirical analysis, and taxonomization. In contrast, other papers use only one or a few of these methodologies to study their subjects;
    \item our paper examines a series of aspects of yield farming, including related DeFi protocols, yield generation, yield farming strategies, financial risks, and security issues, while most of the related papers only cover some of these topics.
\end{itemize}

We discuss related literature in more details from different perspectives in the following subsections.


\subsection{Yield farming}

A few papers focus on studying and comparing yield farming strategies or yield aggregators~\cite{augustin2022yield}. For instance, Nathan Walton~\cite{Walton2022} provides a break down of yield generating mechanism, covering four different farming strategies and a few other related topics (e.g., benefits and risks); Kanis Saengchote~\cite{Saengchote2021stablecoins} studies DeFi composability, which covers yield-chasing behaviors and some introductions about major yield aggregators; another case study about Compound~\cite{Saengchote2022compound} also comes with explanations about yield aggregators and yield farming incentives; Popescu et al.~\cite{Dragos2020} discuss the transition from the traditional finance to DeFi and include some descriptions about yield farming.

However, none of these works have comprehensively investigated yield farming from the perspective of literature survey, modeling, empirical analysis, and taxonomization like our paper.

\subsection{DeFi platforms}

As a type of DeFi application, yield farming is built upon DeFi platforms.
Combing the design of general DeFi platforms lays the groundwork for yield farming designs.
Moin et al.~\cite{moin2019stablecoin} and Pernice et al.~\cite{Pernice2019a} systematically study the general designs of DeFi platforms by decomposing the structure into diverse elements (i.e., peg assets, collateral amount, price and governance mechanism). They also investigate the merits and demerits of DeFi platforms to spot future directions. Nonetheless, yield farming is not the main focus of these works.

\subsection{Related DeFi protocols}

There are various papers studying the DeFi protocols (e.g., flash loan, lending, trading) that can be leveraged by yield farming.
For example, as fundamental protocols of yield farming, the mechanisms, properties, and risks of DeFi lending protocols are extensively investigated in several publications~\cite{Bartoletti2020sokLendingPools,Perez2020liquidations,Tolmach2021,Kao2020a}; some papers~\cite{Gudgeon2020PLF,Perez2020liquidations} provide analysis and discussions about protocols for Loanable Funds, introducing the interest rate determination and liquidity issues; Han et al.~\cite{han2022} zoom into the launch event of the yield farming protocols for Uniswap liquidity provision and further establish the causal impact of this on Binance investor trading activities; within the scope of the analysis of financial attack vectors that involve a flash loan, Qin et al.~\cite{Qin2020c} study the existing flash loan-based attacks and propose optimizations that significantly improve the ROI of these attacks; Gudgeon et al.~\cite{Gudgeon2020c} explore how design weaknesses in \ac{defi} protocols can trigger a decentralized financial crisis.

Although these papers can cover almost all the \ac{defi} protocols used by yield farming, our paper presents this topic more systematically by putting together all the relevant protocols, components, and problems worth exploring.

\section{Conclusion}
\label{sec:conclusion}

In this survey paper, we examine yield farming protocols from multiple perspectives. We first highlight yield farming's dependence on lower-level \ac{defi} primitives in the context of the broader \ac{defi} ecosystem and propose a general framework for yield farming protocols. We then explain code-level actions and associated yield farming operations. We decompose various aspects of yield farming protocols such as protocol form, pool structure, accepted token types, and enumerate their variations.
Later, we stylize three frequently used strategies and simulate yield farming performance under a set of assumptions.
We also compare four major yield aggregators by summarizing their strategies and revenue models.
Finally, we discuss security and economic risks of yield farming protocols, together with related work.

While yield farming has been exploding since 2020, an important question remains if current yields will be sustainable in the long term. Higher rewards also imply higher risks, and associated \ac{defi} attacks prove that the safest and most robust yield provider will win the race. 
Besides security enhancement, new industry developments should consider building one-stop-shop solutions, in pursuit of aggregating more than just yield and facilitating the on-boarding of new \ac{defi} users. 


\section*{Acknowledgment}

We would like to extend our sincere thanks to Simon Cousaert and Toshiko Matsui for the base of this paper \cite{Cousaert2021}.
We also would like to thank Yitian Wang and Honglin Fu for their research assistance.

This material is based upon work partially supported by Ripple
under the University Blockchain Research Initiative (UBRI)~\cite{9935809}. Any
opinions, findings, and conclusions or recommendations expressed
in this material are those of the authors and do not necessarily
reflect the views of Ripple.
\printacronyms

\bibliographystyle{IEEETranS}
\bibliography{references_update}
\begin{IEEEbiography}[{\includegraphics[width=1in,height=1.25in,clip,keepaspectratio]{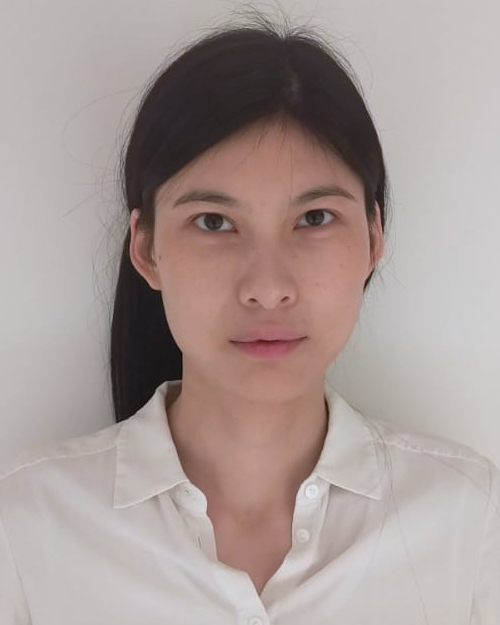}}]{Jiahua Xu}
    is a Lecturer in Financial Computing, and Programme Director of the MSc Emerging Digital Technologies at the Computer Science Department of UCL.
    She is also affiliated to the UCL Centre for Blockchain Technologies, and a founding member of the UK FinTech Academic Network.
    Her research focuses on blockchain economics and decentralized finance.
    She has published in Usenix Security, ACM IMC, FC, IEEE ICDCS and IEEE ICBC. 
    She has reviewed for Advances in Complex Systems, Computer Networks, Transactions on the Web and Cities.
\end{IEEEbiography}

\begin{IEEEbiography}
[{\includegraphics[width=1in,height=1.25in,clip,keepaspectratio]{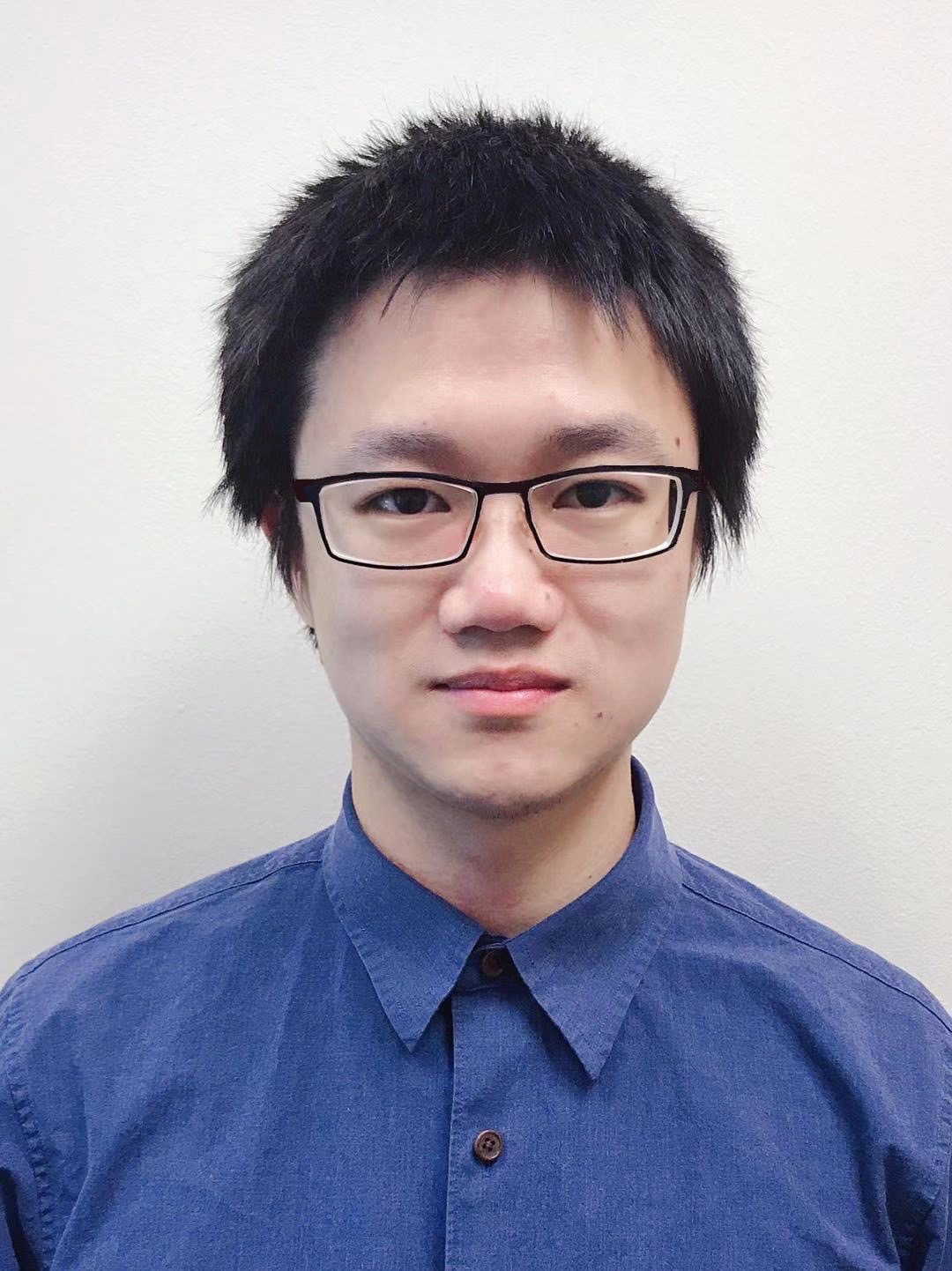}}]{Yebo Feng}
    is a Ph.D. candidate in the Department of Computer and Information Science at the University of Oregon (UO),
    where he conducts his research in the Center for Cyber Security and Privacy.
    His research interests include network security, blockchain security, and network traffic analysis.
    He is the recipient of the Best Paper Award of 2019 IEEE CNS, Gurdeep Pall Graduate Student Fellowship of UO,
    and Ripple Research Fellowship.
    He has served as the reviewer of IEEE TDSC, IEEE TIFS, IEEE JSAC, and ACM TKDD.
\end{IEEEbiography}

\newpage

\appendix
\subsection{Simulating yield Farming Strategies}
\label{sec:formalization}

A yield farming strategy is made of a specific set of actions (see \autoref{sec:protocol-actions}) through modular smart contracts that automates the yield farming process. 
In this section, we describe three common yield farming strategies: {\it simple lending},
{\it leveraged borrow},
and {\it liquidity provision}.
\revision{Besides, to intuitively and roughly compare the performance differences of these yield farming strategies, we simulate each strategy in a controlled, parameterized, simplified environment by tracking the trajectory of the total value $W$ of the yield aggregator\footnote{The code repository can be found from this URL: \href{https://github.com/xujiahuayz/yieldAggregators}{https://github.com/xujiahuayz/yieldAggregators}.}.}

\subsubsection{Assumptions}

\begin{figure}[ht]
\centering
\includegraphics[width=0.7\linewidth]{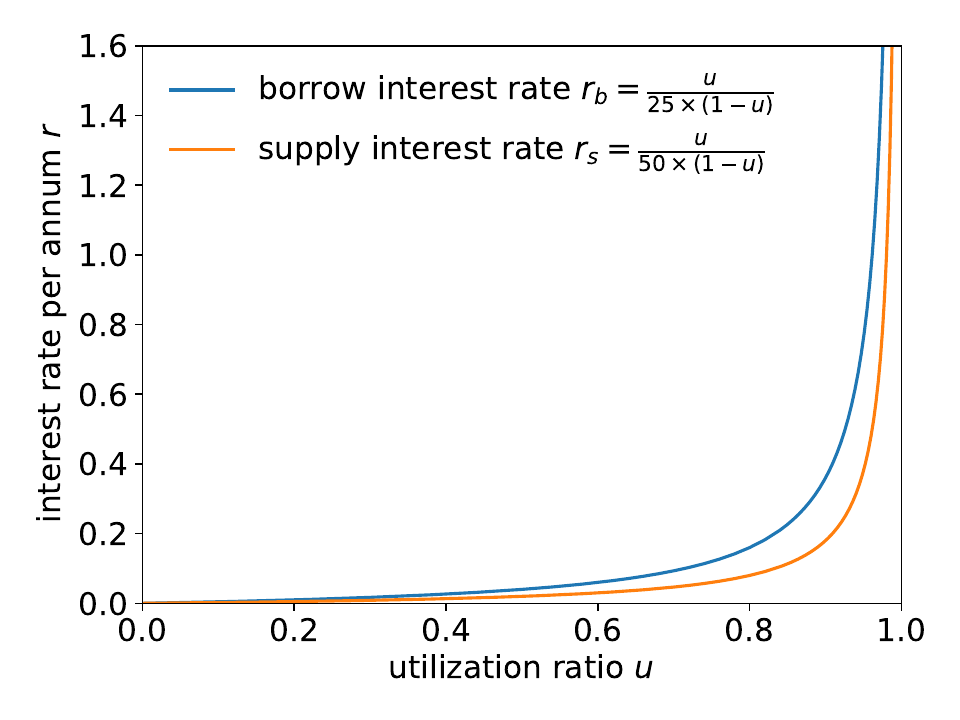}
\caption{Interest rate model for the simulation.
\label{fig:interest}
}
\end{figure}

On comparing the three common strategies, for simple demonstration purposes without loss of generality, we make the following assumptions:

\begin{enumerate}[label=\arabic*.]
    \item the transaction cost is neglected;
    \item the value of the yield farming pool $W_t$ is measured in \coin{USDT}; at $t = 0$, the pool contains 1 \coin{USDT}'s worth of funds, i.e., $W_0 = 1$;
    \label{ass:farmingpoolvalue}
    \item the pool initially supplies all its funds to a yield-generating protocol---either a \ac{plf} or an \ac{amm}, and the funds represent 1\% of the protocol's total assets held at $t = 0$;
    \label{ass:2}
    \item the yield-generating protocol---either a \ac{plf} or an \ac{amm}---distributes 0.01 governance token per day to its users proportionately to their stake in the protocol:
        \begin{enumerate}[label=\alph*., ref=\theenumi{}\alph*]
            \item for a \ac{plf}, half of the governance tokens are distributed to lenders proportionate to their deposits, and half to borrowers proportionate to their loans,
            \label{item:plfgov}
        \item for an \ac{amm}, the governance tokens are distributed proportionately to \acp{lp};
        \end{enumerate}
    \item the governance token price remains constant during the simulation period;
    \item the lending platform has an non-linear interest rate model \cite{Gudgeon2020PLF} as illustrated in \autoref{fig:interest};
    \label{item:borrowapy}
    \item the \ac{amm} has a fixed exchange fee of 5\% and applies a Uniswap-like constant-product conservation function;\footnote{We refer the reader to \cite{Angeris2020c} for detailed description and analyses on \acp{amm} with constant-product conservation function.} the fee is charged by retaining 5\% of the theoretical fee-free purchase quantity within the \ac{amm} pool.
    \label{ass:amm}
\end{enumerate}

\subsubsection{Simulation results}
\label{sec:sim-results}

\paragraph{Simple lending}
\label{sec:simplelending}

The yield farming pool grows its wealth through accrual of supply interest and reward tokens distributed by the \ac{plf}.

\subparagraph{Simulated strategy execution}
In our simulated environment, at $t=0$ the yield aggregator deposits 1 \coin{USDT} to a \ac{plf}, and receives in return some \coin{interest-USDT} as a certificate of deposit (see \autoref{sec:plf}). According to Assumption~\ref{ass:2}, the aggregator owns 1\% of the total circulating supply of \coin{interest-USDT}. 

The \coin{interest-USDT} holding of the aggregator is worth exactly 1 \coin{USDT} at $t=0$, and increases in value due to interest accrued with the passage of time. In addition, the farming pool is rewarded with the \ac{plf}'s governance tokens owing to its supply contribution, and the value of the governance token holding is counted towards the total value of the yield farming pool.

\subparagraph{Simulated scenarios}
At each given rewarded protocol token price, we simulate three scenarios: the initial utilization ratio of the funds in the lending pool equals 0, 0.4, 0.8, respectively. As illustrated in \autoref{fig:interest}, a higher utilization ratio indicates a higher supply interest rate.

\begin{figure}[tp]
\centering
\subfloat[Simple lending.\label{fig:simple_lending_sim}]
    {\includegraphics[height=0.103\textheight, trim = {13, 0, 11, 0}, clip]{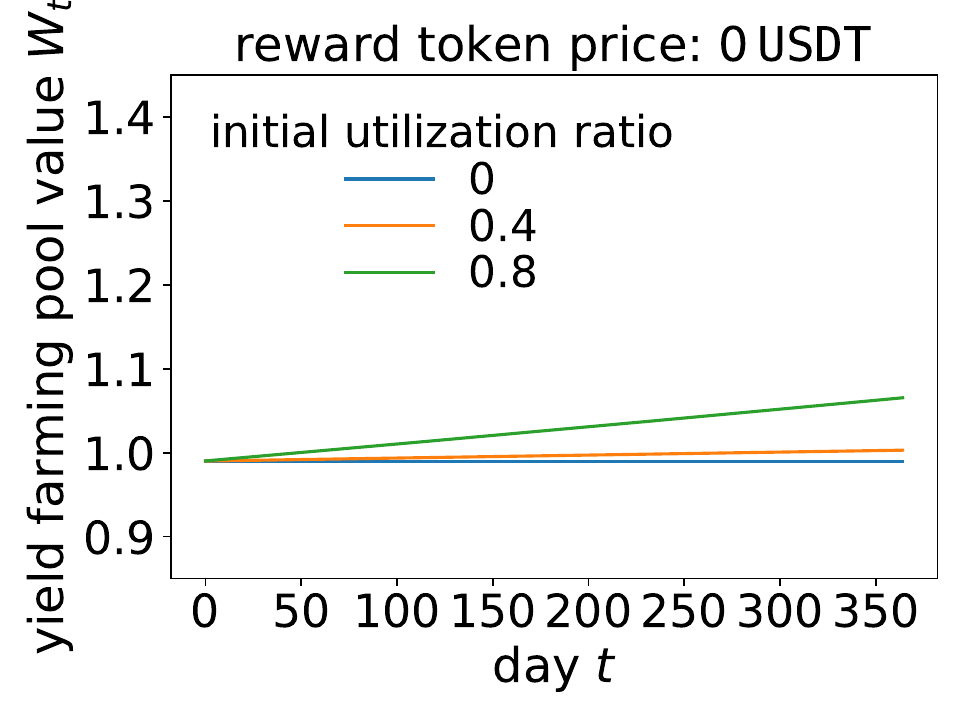}
    \includegraphics[height=0.103\textheight, trim = {82, 0, 11, 0}, clip]{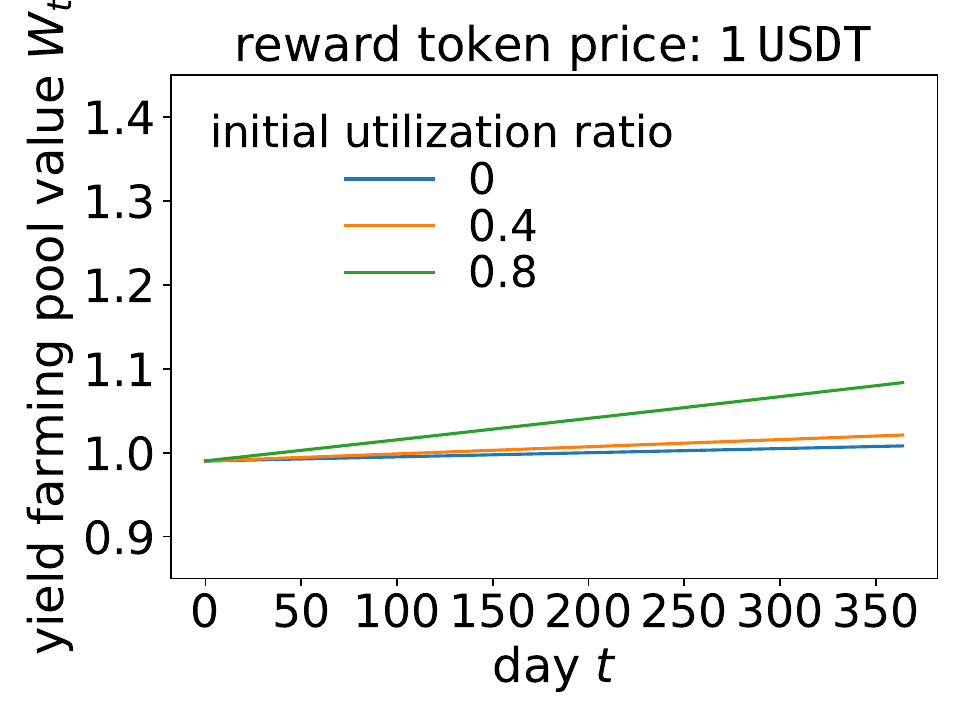}
    \includegraphics[height=0.103\textheight, trim = {82, 0, 11, 0}, clip]{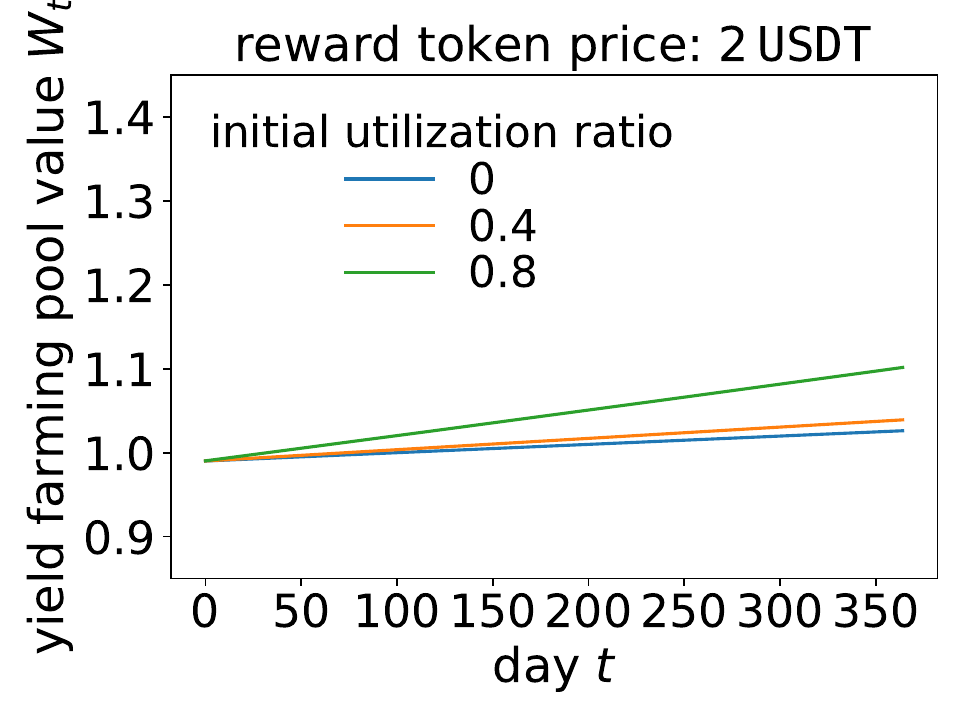}}

\subfloat[Leveraged borrow.\label{fig:spiral_lending_sim}]
    {\includegraphics[height=0.103\textheight, trim = {13, 0, 11, 0}, clip]{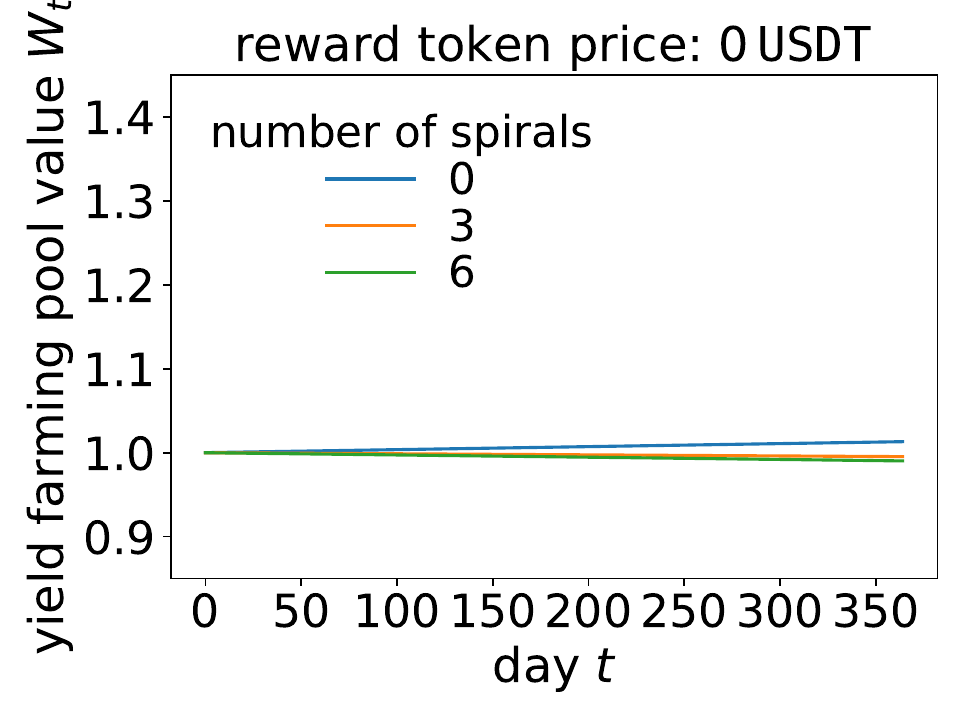}
    \includegraphics[height=0.103\textheight, trim = {82, 0, 11, 0}, clip]{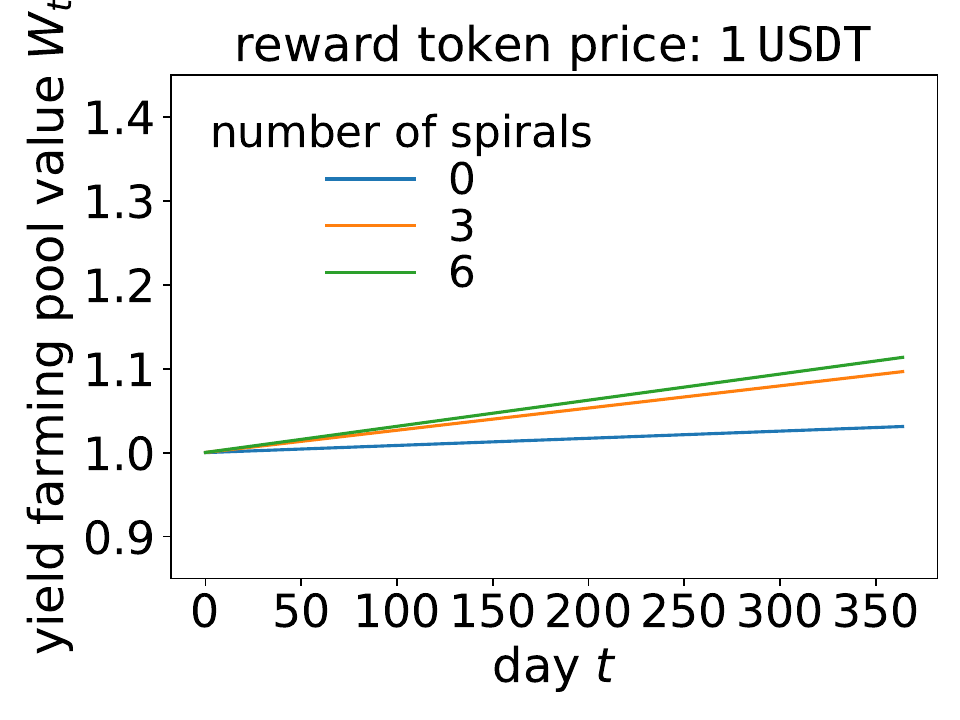}
    \includegraphics[height=0.103\textheight, trim = {82, 0, 11, 0}, clip]{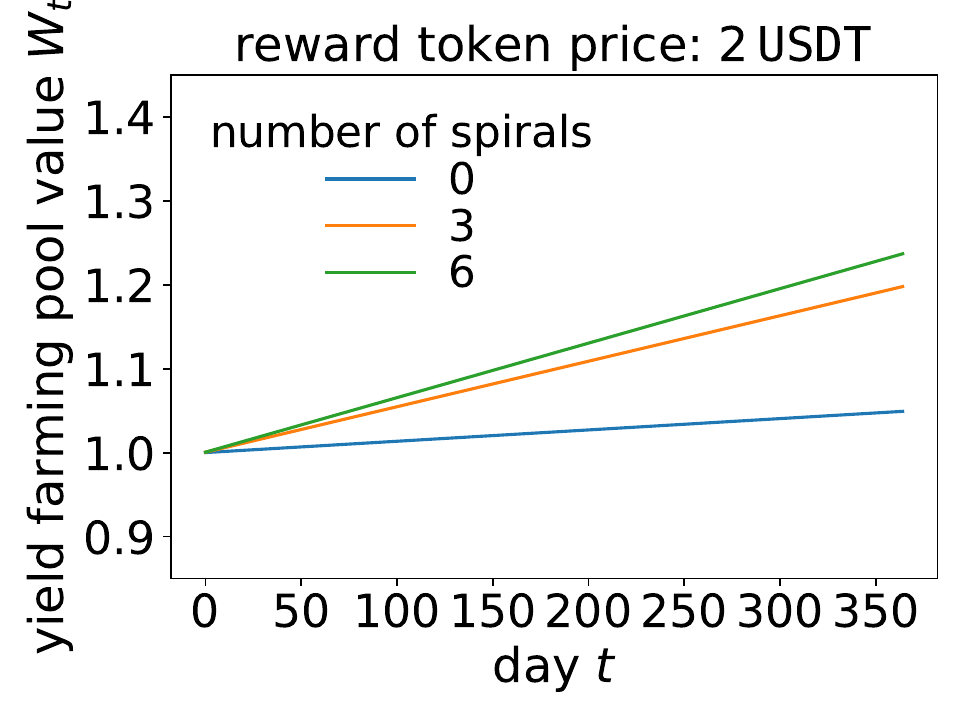}}

\subfloat[Liquidity provision.\label{fig:lp_mining_sim}]
    {\includegraphics[height=0.103\textheight, trim = {13, 0, 11, 0}, clip]{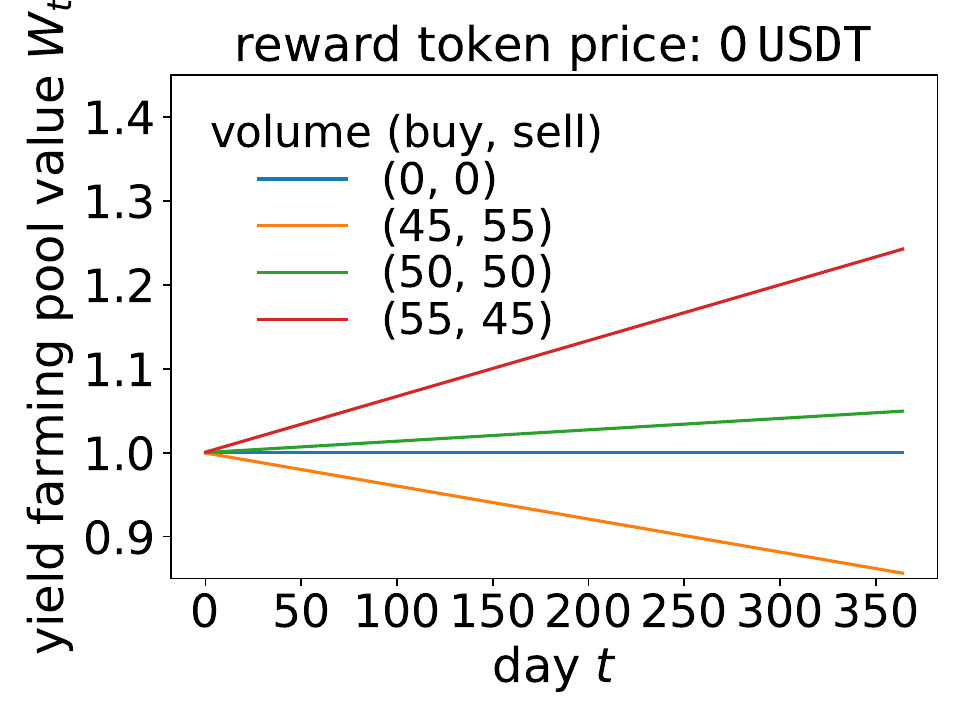}
    \includegraphics[height=0.103\textheight, trim = {82, 0, 11, 0}, clip]{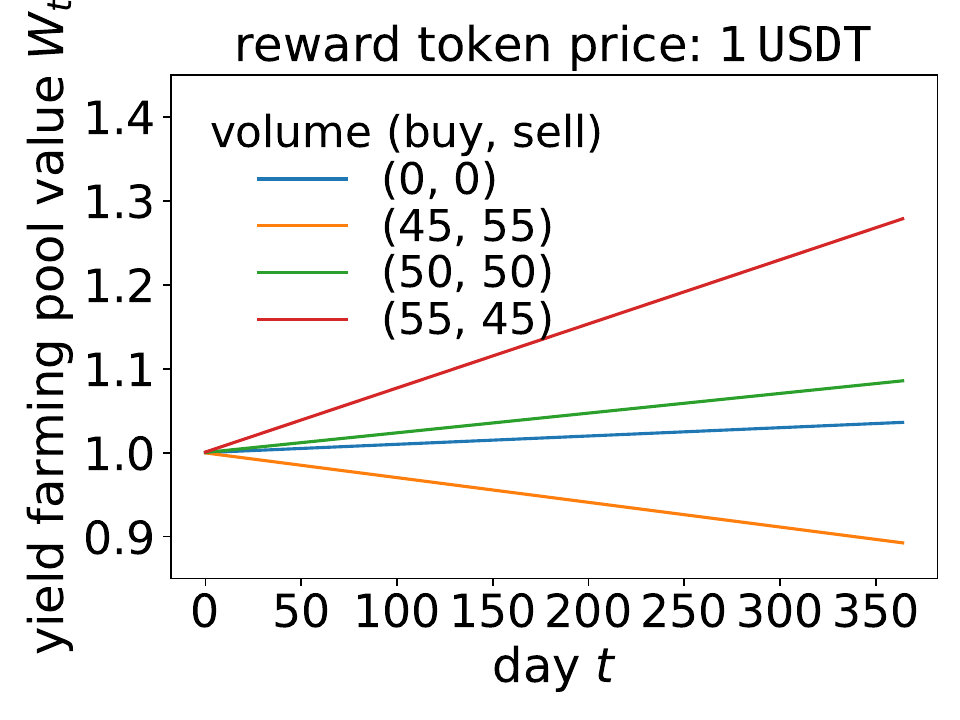}
    \includegraphics[height=0.103\textheight, trim = {82, 0, 11, 0}, clip]{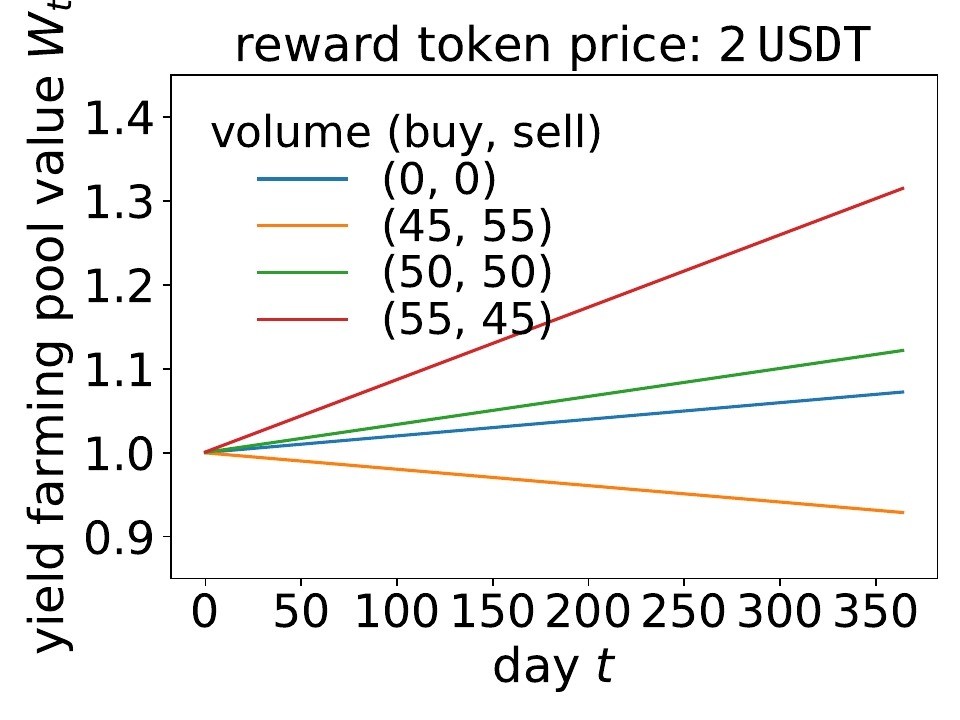}}

\caption{Simulation results of various yield farming strategies.}
\label{fig:simulation_results}
\end{figure}

\subparagraph{Results}
Figure~\ref{fig:simple_lending_sim} shows that, 
the value held by the aggregator $W_t$ is floored at 1 \coin{USDT}, with the worst-case scenario when the supply interest equals 0 due to the absence of borrow demand and the reward token has 0 value. Intuitively, $W_t$ increases with higher utilization and reward token price.

\paragraph{Leveraged borrow}
\label{sec:leveragedborrow}

According to Assumption~\ref{item:plfgov} and in line with practices of major lending platforms such as Compound \cite{Compound2021}, governance tokens are rewarded to both lenders and borrowers. This strategy thus aims to maximize the amount of governance tokens received by the lending platform through leveraging spirals. 

\subparagraph{Simulated strategy execution}
In our simulated environment, at $t=0$ the yield aggregator first deposits 1 \coin{USDT} to a lending platform;
with this initial deposit as collateral, the aggregator then takes a loan worth 65\% of its deposit, i.e. 0.65 \coin{USDT}. 
To further augment its deposit and borrow amount for the entitlement of larger rewards, the aggregator re-deposits the borrowed funds, and use them as collateral to borrow again 0.65\% of the new deposit; and so on and so forth. Obviously, the more spirals the yield farming pool undertakes, the higher shares it holds at both the lending and the borrowing sides of the lending platform.   

\subparagraph{Simulated scenarios}
We assume the initial utilization ratio of the \ac{plf}'s lending pool is 0.4. 
At each given reward token price, we simulate three scenarios: 
\begin{enumerate*}[label={(\roman*)}]
\item depositing without borrowing,
\item lending and repeat borrowing and re-supplying 3 times,
\item lending and repeat borrowing and re-supplying 6 times.
\end{enumerate*}

\subparagraph{Results}
As an asset's borrow interest rate always exceeds its supply interest rate, the loan accrues interest exponentially faster than its deposit. We observe from Figure~\ref{fig:spiral_lending_sim} that sufficiently valuable reward tokens can make the strategy profitable, but losses occur when the value of the governance tokens received is insufficient to offset the negative net interest revenue. 
Overall, a high degree of leverage, measured by the number of spirals, can amplify both the profit---in case of high-value governance tokens, as well as the loss---in case of low-value governance tokens.

\paragraph{Liquidity provision}
\label{sec:liquidityprovision}
The yield farming pool supplies funds to an \ac{amm} in order to profit from both trading fees and governance tokens rewarded by the \ac{amm}.

\subparagraph{Simulated strategy execution}
In our simulated environment, at $t=0$ the aggregator 
deposits 1 \coin{USDT}'s worth of funds in the \coin{USDT-ETH} pool of an \ac{amm}, and receives in return some \coin{USDT-ETH-LP} tokens, representing its share in the \ac{amm} pool. According to Assumption~\ref{ass:2}, the aggregator owns 1\% of the total circulating supply of \coin{USDT-ETH-LP}. Given that \coin{USDT} is the denominating asset (Assumption~\ref{ass:farmingpoolvalue}), and that the \coin{USDT-ETH} pool applies a constant-product conservation function (Assumption~\ref{ass:amm}), the \coin{USDT-ETH} pool always contains \coin{USDT} and \coin{ETH} with equivalent value, and the total pool value thus equals twice the \coin{USDT} quantity in the pool.\footnote{We refer the reader to \cite{xu2021dexAmm} for a formal derivation on the pool value of a constant-product \ac{amm}.}

We additionally assume that, on an aggregate level, further liquidity provision and withdrawal cancel each other out. Hence, the aggregator's ownership of the \ac{amm} pool is neither diluted nor concentrated; that is, the value of the aggregator's \coin{USDT-ETH-LP} holding remains 1\% of the \coin{USDT-ETH} pool's value. Naturally, all other things equal, the value held by the aggregator increases with the value of the \ac{amm} governance token.

\subparagraph{Simulated scenarios}
We test scenarios with different market movements. Specifically, we illustrate in \autoref{fig:lp_mining_sim} when during the entire simulation period
\begin{enumerate*}[label={(\roman*)}]
\item there is 0 trading volume (blue line), 
\label{item:0trading}
\item the buy and sell volume of \coin{ETH} is respectively 45 \coin{USDT} and 55 \coin{USDT} (orange line), 
\label{item:moreselleth}
\item the buy and sell volume of \coin{ETH} is each 50 \coin{USDT} (green line),
\label{item:5050trading}
\item the buy and sell volume of \coin{ETH} is respectively 55 \coin{USDT} and 45 \coin{USDT} (orange line). 
\label{item:morebuyeth}
\end{enumerate*}
We assume that the trading volume is evenly spread out throughout the simulation period.

Absent any trading activity---as in Scenario \ref{item:0trading}, the aggregator's yield solely comes from governance token reward.
The yield difference between Scenarios \ref{item:0trading} and \ref{item:5050trading} lies in the trading fee. By comparing the blue line and the green line in Figure~\ref{fig:lp_mining_sim}, we clearly see that \ref{item:5050trading} results in higher yield with the presence of 5\% trading fee.

Scenario \ref{item:moreselleth} describes a market situation with higher selling pressure and consequently falling \coin{ETH} prices. The leads to an increase in the quantity of the depreciated \coin{ETH} and a decrease in the quantity of the denominating asset \coin{USDT} in the \ac{amm} pool, diminishing the \ac{amm} pool's value. When the trading fee revenue and governance token reward are insufficient to offset this value loss, the yield would be negative. 

In contrast to Scenario \ref{item:moreselleth}, Scenario \ref{item:morebuyeth} describes an opposite market situation where a higher demand in \coin{ETH} drives up its price. The leads to a decrease in the quantity of the appreciated \coin{ETH} and an increase in the quantity of the denominating asset \coin{USDT} in the \ac{amm} pool, raising the \ac{amm} pool's value.

Note that in both Scenarios \ref{item:moreselleth} and \ref{item:morebuyeth}, liquidity providers suffer from divergence loss \cite{xu2021dexAmm}, that is, they could have been better off by just holding their \coin{USDT} and \coin{ETH}, as opposed to supplying them to the \ac{amm} pool.

\subparagraph{Results}
The simulation shows that the liquidity provision strategy also entails risk, associated with market movements of the assets within the \ac{amm} pool. Higher volatility of the \ac{amm} pool assets implies higher uncertainty in yield.

\subsection{Current Major yield aggregators in DeFi}
\label{sec:comparison}

As listed in \autoref{tab:marketshare}, to date, yield aggregators have collected billions of dollars worth of liquidity. This section compares current major yield aggregators with a focus on their strategies, performances, and fee mechanisms. \autoref{tab:top20yieldagg} lists the characteristics of current top 20 yield aggregators. \revision{All the data was collected on 28 August 2022.}

\subsubsection{Yearn Finance}
\label{subsec:yearn}

Yearn Finance offers a multitude of products in \revision{DeFi}, providing lending aggregation, yield generation and others \cite{yearn2021intro}. The services discussed here are Yearn Earn, a lending aggregator, and Yearn Vaults, a more comprehensive yield aggregator. Yearn Finance launched in July 2020.



\paragraph{Strategies}

\begin{itemize}
    \item Earn pools: The strategy of the Earn pools is to collect a certain asset and deposit it either in dYdX, Aave or Compound, depending on where the highest interest rate of that asset is found. Yearn will withdraw from one protocol and deposit to another automatically as interest rates change between protocols, in a strategy that is slightly similar to the Idle Finance ``Best-Yield'' Strategy. Proportional shares in Earn pools are commonly represented by \coin{yTokens}.
    \item Vaults: A Yearn Vault uses an asset as liquidity, deposits that liquidity as collateral (accounting for risk levels) to borrow stablecoins. Then, it uses those stablecoins to generate yield, after which that yield is re-invested in the stablecoins to generate more yield. Vaults thus allow for more complex strategies compared to Earn pools. Proportional shares in Yearn Vaults are commonly represented by \coin{yvTokens} or other \coin{yTokens}.
    
    
\end{itemize}

Yield farming strategies in Yearn v2 Vaults can be complex, involving flash loans (uncollateralized loans that are taken and repaid within the same transaction \cite{Xu2021c}), leveraged borrowing, staking on specific protocols (for example HegicStaking) and more.

\paragraph{Return for users}

Yearn Finance distributed the \coin{YFI} governance token over a 9-day period after launch. Liquidity providers in the Earn pools or Vaults are thus not incentivized by a Yearn liquidity mining program, so current yield only comes from the returns that the product strategies reap. Those returns can be straightforward, as is the case for Earn pools, and can be complex to calculate, as is the case for v2 Vaults that can have up to 20 strategies working at once. Some Yearn vaults accept LP tokens, other accept single asset tokens.

\paragraph{Protocol fees}

v1 Vaults have a 20\% performance fee and a 0.5\% withdrawal fee (in case funds need to be pulled from the strategy in order to cover the withdrawal request). v2 Vaults have also a 20\% performance fee, but no withdrawal fee. Instead, they charge a 2\% management fee. Performance fees are split 50:50 between the Treasury and the Strategist, the official creator of the strategy. The management fee is assigned fully to the Treasury.

\subsubsection{Idle Finance}

Launched in August 2019, Idle is a yield aggregator that automatically allocates and aggregates interest-bearing tokens \cite{idle2021documentation}. 



\paragraph{Strategies}

Idle Finance only distributes single-asset pools over different lending protocols.
Users' funds are pooled together and depending on the strategy that the pool employs, assets are allocated over different lending platforms, currently limited to: Compound, Fulcrum, Aave, DyDx and Maker DSR. In Idle Finance, the currently supported pools can be deposited directly into the above lending platforms. When any user interacts with Idle or if no interactions are made for 1 hour, rebalancing of the assets takes place according to the rates of supported providers.

Currently, Idle uses two different allocation strategies:
\begin{itemize}
    \item Best-Yield: this strategy seeks the best interest rates across multiple lending protocols.
    \item Risk-Adjusted: this strategy automatically changes the asset pool allocation in order to find an allocation with the highest risk-return score, compared to the highest return score of the \enquote{Best Yield} strategy. It does this by incorporating a framework for quantifying risk, developed by DeFiScore \cite{defiscore2021}, which outputs a 0-10 score that represents the level of risk on a specific lending platform (0 = highest risk, 10 = lowest risk).
\end{itemize}

\paragraph{Return for users}

Idle uses {\tt IdleTokens} to represent the farmers' proportional ownership of the asset pool, which should accrue yield over time. In addition, farmers are rewarded with \coin{IDLE} governance tokens for participating in the pools as part of Idle's liquidity mining program. In January 2021, a two-year liquidity mining program started to reward liquidity providers depending on the amount of funds deposited and the utility generated by a certain pool \cite{idle2021distribution}.

\paragraph{Protocol fees}

A performance fee 10\% of the generated yield is charged.


\subsubsection{Harvest Finance}

Harvest Finance gives \coin{FARM} holders the opportunity to share in the revenue model of the protocol. By staking \coin{FARM}, users are entitled to receive part of the revenue that is collected by the protocol. Harvest Finance went live in August 2020, and currently has more than 
70 pools/vaults in its offering. 



\paragraph{Strategies}

Harvest Finance has two main categories of yield farming strategies~\cite{harvest2021strategies}:

\begin{itemize}
    \item Simple single-asset Strategies: Users deposit single assets such as \coin{USDC}, \coin{USDT}, \coin{DAI}, \coin{WBTC}, \coin{renBTC} or \coin{WETH} into a Harvest Vault, which then deposits those assets into another yield generating protocol, including Compound and Idle Finance. 
    \item LP token Strategies: Users deposit LP tokens from Uniswap, Sushiswap or Curve into Harvest which automatically collects liquidity mining rewards and re-invests them into LP tokens.
\end{itemize}

\paragraph{Return for users}

Depending on the vault used, return of Harvest users is composed of
\begin{enumerate*}[label={(\roman*)}]
\item the fees accrued by providing liquidity to AMM pools or other yield-bearing assets, 
\item earning tokens distributed through external liquidity mining programs and 
\item extra \coin{FARM} tokens as part of the liquidity mining program. 
\end{enumerate*}
These returns are dependent on underlying market forces, liquidity programs and token values. For example, the Harvest emission schedule defines how much \coin{FARM} will be distributed over time \cite{harvest2021token}.


\paragraph{Protocol fees}

Harvest Finance does not charge withdrawal fees and does not claim a direct ``fee'' on the yield farming revenue. However, during liquidation of the yield, 30\% of the profits is used to buy the \coin{FARM} token on the market, which is then distributed to users who stake \coin{FARM} in the profit-sharing \coin{FARM} pool \cite{harvest2021vaultFunction}.\footnote{This type of buyback reduces the supply of governance tokens in the secondary market to the benefit of existing tokenholders.}

\subsubsection{Pickle Finance}

Launched in September 2020, Pickle offers yield on deposits through two products: Pickle Jars (pJar) and Pickle Farms. Jars are yield farming robots, earning returns on users' funds, while farms are liquidity mining pools where users can earn {\tt PICKLE} governance tokens by staking different kinds of assets. Proportional shares in pJars are represented by \coin{pTokens}.



\paragraph{Strategies}

Each Pickle Jar employs a specific strategy to earn yield. Currently, two main versions of pJar strategy are in existence, pJar 0.00 and pJar 0.99, of which the 0.99 version is most important. In either version, pooled funds are directly utilized to farm rewards, after which they are sold to re-invest the accrued yield.

\begin{itemize}
    \item pJar 0.00: These pJars involve a user depositing LP tokens acquired by supplying liquidity on Curve Finance \cite{Curve2021}, an AMM-based DEX. The strategy employed in pJar 0.00 earns and re-invests \coin{CRV} rewards by selling \coin{CRV} into the market for stablecoins and re-depositing those into the Curve pools to get more LP tokens. Effectively, pJars 0.00 generate yield by accruing
    \begin{enumerate*}[label={(\roman*)}]
    \item LP fees from Curve and
    \item generating \coin{CRV} tokens because of Curve's liquidity mining program \cite{curve2021tokenomics}.
    \end{enumerate*}
    \item pJar 0.99: These pJars utilize LP tokens from Uniswap and Sushiswap, earning yield by accruing
    \begin{enumerate*}[label={(\roman*)}]
    \item LP fees from Uniswap/ Sushiswap and 
    \item generating \coin{SUSHI} or other native tokens because of liquidity mining programs.
    \end{enumerate*}
\end{itemize}

\paragraph{Return for users}

Return of Pickle users is generally composed of
    \begin{enumerate*}[label={(\roman*)}]
    \item the fees accrued by providing liquidity to AMM pools,
    \item earning tokens distributed through external liquidity mining programs, and
    \item extra \coin{PICKLE} tokens if the yield farmer makes use of the Farm products. 
    \end{enumerate*}
    The return is thus dependent on underlying market forces, liquidity programs and token values. For example, the Pickle emission schedule defines how much \coin{PICKLE} will be distributed over time \cite{pickle2021emission}. 


\paragraph{Protocol fees}

Most Pickle Jars have a 20\% performance fee on the generated yield.


\subsubsection{Other aggregators}

The four main yield aggregators above are deemed the most mature, but a new wave of yield aggregating protocols is coming up. In general, there seems to be a tendency where more recent yield aggregators aim to be a one-stop-shop, providing additional functionalities such decentralized exchanges, lending and borrowing and risk-managing services. This enhances user experience and introduces more revenue streams for the protocols.

Below we list more recently launched protocols and products that are still being tested.

\paragraph{Rari Capital}
Rari Capital \cite{rari2021} is a roboadvisor that attempts to provide investors with the highest yield, beyond just lending. It has multiple products, including Earn, Tranches, Fuse and Tanks. The Earn product can be considered a traditional yield farming service, while the other products are extending the number of functionalities on the Rari Capital platform, such as lending and borrowing, and yield farming within certain risk boundaries, called ``tranches'' \cite{rariTranches2020}.

\paragraph{Vesper Finance}
Vesper \cite{vesper2021} 
focuses on institutional adoption of the DeFi yield market. Currently, only Vesper Grow Pools are available, which are comparable to the traditional yield products. In future developments, Vesper plans to integrate 
Vesper Labs \cite{vesper2021} where external users can build their own strategy in return for part of the reaped profits. 

\subsubsection{Summary}

Many yield farming strategies entail some extent of optimization, e.g. choosing the lending pool that offers the highest APY to deposit assets into (e.g. Idle's Best-Yield pools, Yearn's earn pools), or balancing between risks and return (e.g. Idle's risk-adjusted pools). The core strategies applied by major yield aggregators commonly do not deviate much from the basic strategies described in \autoref{sec:formalization}. However, as the competition in yield farming grows, basic strategies becomes less effective \cite{Jakub}, which prompts protocols to device more sophisticated strategies that incorporate various forms of interactions with other DeFi protocols (e.g. upgrade from Yearn v1 to v2). Yield aggregators generate revenues by charging fees from investors. Protocols associated with better yield farming performance are able to charge higher fees, which can be observed by comparing both the performance fee between Yearn (20\%) and Idle (10\%) as well as their respective performance.


\end{document}